# Seismic waves in a three-dimensional block medium

## N.I. Aleksandrova

*N.A. Chinakal Institute of Mining, Siberian Branch, Russian Academy of Sciences,*

*Krasnyi pr. 54, Novosibirsk 630091, Russia*

e-mail: *nialex@misd.ru*

**Abstract.** We study numerically the propagation of seismic waves in a three-dimensional block medium. The medium is modeled by a spatial lattice of masses connected by elastic springs and viscous dampers. We study Lamb's problem under a surface point vertical load. The cases of both step and pulse load are considered. The displacements and velocities are calculated for surface masses. The influence of the viscosity of the dampers on the attenuation of perturbations is studied. We compare our numerical results for the block medium with known analytical solutions for the elastic medium.

**1. Introduction**

Until recently, geomechanics and geophysics widely used the theory of deformation of rock masses which are treated as homogeneous media. In this theory, dynamical properties of the medium are described by the well-developed linear theory of elastic waves. On the basis of this theory, methods are developed for calculating the stress state of rocks near the mines as well as for processing and interpretation of seismological data in geophysics and mining. Serious reason to revise established views give the papers [1–3], which show that, in the mathematical models of rock masses, it is necessary to take into account the block-hierarchical structure of the medium. Especially strong influence on the development of such an approach has been made by a fundamental concept proposed by Sadovskiy [1], according to which, rock masses are treated as systems of nested blocks of different scales, connected with each other by interlayers composed of weaker fractured rocks. Such a structure of the medium affects the process of wave propagation. The presence of the interlayers with weakened mechanical properties leads to the fact that the deformation of the block medium is mainly due to the deformation of the interlayers. As noted in [2], the structure of a block medium is the cause of various dynamic phenomena that are absent in a homogeneous medium and, therefore, cannot be described by its models. Among those dynamic phenomena we distinguish the propagation of the pendulum waves [2,3] having a low velocity of propagation (that is significantly less than the velocity of elastic waves in the blocks), long length (even under a pulse action) and weak damping.

In [4–9], we studied one-dimensional mathematical models of viscoelastic deformation of block media and have shown that the presentation of the blocks as massive rigid bodies allows us to distinguish from a complex dynamic state of the block medium that part of it, which is determined by the deformation of the interlayers between the blocks. The presentation of the blocks as massive rigid bodies made it possible to accurately describe the low-frequency pendulum waves arising under the impact load. In [4–9] it is shown that impulse load generates broadband vibrations in a block medium which, with the spread, is divided into high-frequency oscillations, corresponding to the eigen oscillations of the blocks, and low-frequency oscillations. Laboratory experiments on physical models of block media described in [3–7] have shown that high frequency waves attenuate rapidly and the main seismic effects are caused by low frequency waves. In contrast with the studies [2–7], which focuses on the study of the results of laboratory and field experiments in block media, in [8,9], a theoretical analysis is carried out of the propagation and spectral characteristics of waves under the influence of the block-hierarchical structure of the rock mass.

The dynamic behaviour of a two-dimensional block medium was studied in [10–13]. In these papers the same approach has been used that was employed in [9] for a one-dimensional block medium. In [10–13], it was assumed that the rectangular blocks interact with each other via compressible elastic interlayers. In [10], the blocks were assumed to be rigid, while in [11–13] they were assumed to be elastic. In [12,13], the equations of the orthotropic Cosserat continuum were also used for the description of the dynamic properties of a two-dimensional block medium.

A more simple model of a two-dimensional block medium can be obtained if the blocks are treated as point masses connected with each other by springs and dampers. In this case, a regular block medium can be represented as a periodic lattice of the masses. Different versions of this model were used in [14–22] in order to describe the plane and antiplane deformations of discrete media. However, in [22], the emphasis is on continuous models describing discrete microstructures.

Despite the obvious limitations of the usage of periodic lattices as models for the description of real-world block media, they have an advantage of making it possible to use analytical and numerical methods, as well as to describe qualitatively the dynamic effects inherent to these media. Initially the theory of waves in periodic structures appeared in the theory of crystal lattice dynamics [23–25], later it found applications in the mechanics of composites [26], but for a long time it was not used for description of dynamic properties of rock masses. However, the study of wave propagation in periodic block structures is important for applications in seismology [1–3].

Below we draw attention of the reader to some publications devoted to the propagation of seismic waves which are relevant to our study. For the first time, the problem of the dynamic impact of a vertical point force applied onto the surface of an elastic half-space was considered by Lamb [27]. In that paper he showed that, on the surface of the half-space, the Rayleigh waves [28] propagate along with the P- and S-waves. A short review of results on plane Lamb's problem can be found, e.g. in [18]. The spatial transient Lamb's problem and Rayleigh waves were studied in many monographs, e.g., [29,30], and articles (e.g. [20,31–39]).

In [30], Cagniard obtained a solution to the three-dimensional Lamb's problem using the method of integral transformations. To calculate the inverse integral transforms, he proposed a method that now bears his name. Later that method was developed and generalized in [31]. In [32], Ogurtsov studied Lamb's problem for both vertical and horizontal impact onto the surface of an elastic half-space. In [33–35], the authors studied the propagation of seismic waves in an elastic half-space caused by a transient spherically symmetric source embedded into the half-space. In [33], the emphasis is on the theoretical calculations of the displacements on the surface of the half-space. In [35], analytical representations are obtained for the displacements and stresses in the Rayleigh wave on the surface of the homogeneous elastic half-space and P-wave inside it. In [36], the influence of inhomogeneity of the elastic half-space was studied. In [37], the analysis of the Rayleigh waves was conducted in the framework of the Cosserat continuum. In [38], the authors used the finite-element method for plane and spatial Lamb's problems for the case of harmonic loading. In [39], Kausel generalized analytical results obtained for Lamb's problem for an elastic medium for the cases of the vertical load, horizontal load and embedded source. The paper [39] contains many references related to the spatial Lamb's problem for an elastic medium. In [20], a solution is given to Lamb's problem in a discrete medium. That solution was obtained for two-dimensional and three-dimensional lattices in the form of multiple integrals, which are similar to the integral representations of the Bessel functions.

In the present paper, a block medium is modeled as a three-dimensional lattice of masses connected by springs and dampers in the axial and diagonal directions. For this model, we (i) study dispersion properties; (ii) solve Lamb's problem using a finite-difference method; (ii) study the degree of attenuation of numerical solutions to Lamb's problem on the surface of the block medium as a function of the distance from the impact point and the viscosity of the dampers; and (iv) compare that numerical solution with the results of analytical solutions for a homogeneous elastic half-space presented in [39].

**2. Setting of the problem**

We study the transient spatial Lamb's problem on the impact of a vertical point load on the surface of a half-space filled by a block medium.

The block medium is modeled by a uniform three-dimensional lattice consisting of the point masses, connected by springs and dampers in the directions of the axes *x, y, z*, and in the diagonal directions in the planes *xy, xz, yz*, as shown in figure 1. In that figure, *u, v* are the horizontal displacements in the directions of the axes *x*, *y*; *w* is the vertical displacement in the direction of the axe *z*; *n*, *m*, *k* ($k<0$) are the indices of the masses in the directions of the axes *x*, *y*, *z*. Our theoretical description of the deformation of interlayers is based on the rheological model by Kelvin–Voigt [40]. A vertical point load is applied suddenly onto the surface of the block medium ($k=0$) at the point with the coordinates $n=0$, $m=0$ (figure 2).

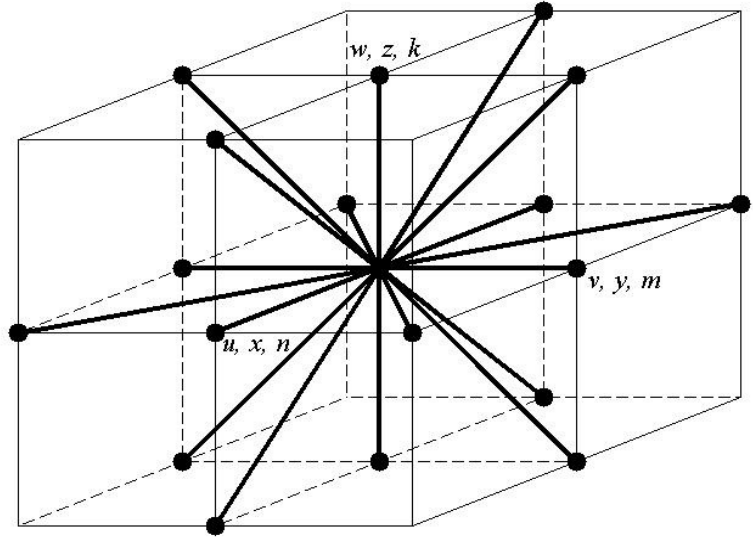

**Figure 1.** Scheme of connections of the masses by springs and dampers in the three-dimensional model of a block medium.

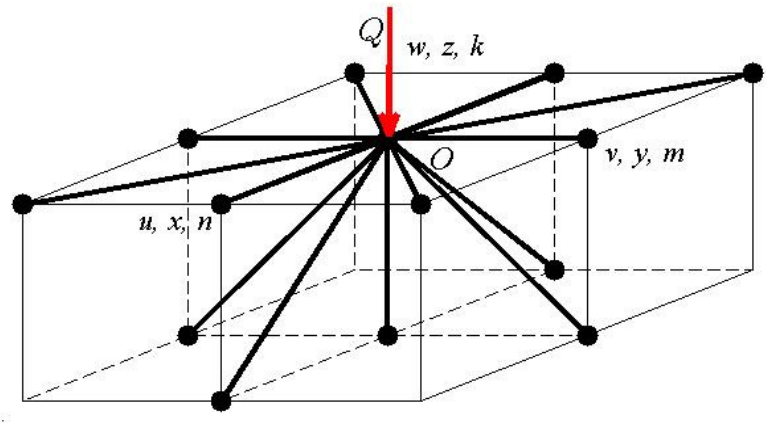

**Figure 2.** Scheme of connections of the masses by springs and dampers on the surface on the half-space in the three-dimensional model of a block medium and the vertical point load.

Using the notation

$$\left.\begin{aligned}\Lambda_{nn}f_{n,m,k} &= f_{n+1,m,k} - 2f_{n,m,k} + f_{n-1,m,k}, \\ \Phi_{nm}f_{n,m,k} &= f_{n+1,m+1,k} + f_{n-1,m-1,k} - 4f_{n,m,k} + f_{n+1,m-1,k} + f_{n-1,m+1,k} \\ \text{and}\quad \Psi_{nm}f_{n,m,k} &= f_{n+1,m+1,k} + f_{n-1,m-1,k} - f_{n+1,m-1,k} - f_{n-1,m+1,k},\end{aligned}\right\} \quad (2.1)$$

we write the equations of the motion of the mass with indices *n*, *m*, *k* ($k<0$), located inside the half-space, in the following form:

$$\left.\begin{aligned}
M\ddot{u}_{n,m,k} &= k_1\Lambda_{nn}u_{n,m,k} + \frac{k_2[(\Phi_{nk}+\Phi_{nm})u_{n,m,k} + \Psi_{nm}v_{n,m,k} + \Psi_{nk}w_{n,m,k}]}{2} \\
&\quad + \lambda_1\Lambda_{nn}\dot{u}_{n,m,k} + \frac{\lambda_2[(\Phi_{nk}+\Phi_{nm})\dot{u}_{n,m,k} + \Psi_{nm}\dot{v}_{n,m,k} + \Psi_{nk}\dot{w}_{n,m,k}]}{2} \\
M\ddot{v}_{n,m,k} &= k_1\Lambda_{mm}v_{n,m,k} + \frac{k_2[\Psi_{nm}u_{n,m,k} + (\Phi_{mk}+\Phi_{nm})v_{n,m,k} + \Psi_{mk}w_{n,m,k}]}{2} \\
&\quad + \lambda_1\Lambda_{mm}\dot{v}_{n,m,k} + \frac{\lambda_2[\Psi_{nm}\dot{u}_{n,m,k} + (\Phi_{mk}+\Phi_{nm})\dot{v}_{n,m,k} + \Psi_{mk}\dot{w}_{n,m,k}]}{2} \\
\text{and}\quad M\ddot{w}_{n,m,k} &= k_1\Lambda_{kk}w_{n,m,k} + \frac{k_2[\Psi_{nk}u_{n,m,k} + \Psi_{mk}v_{n,m,k} + (\Phi_{mk}+\Phi_{nk})w_{n,m,k}]}{2} \\
&\quad + \lambda_1\Lambda_{kk}\dot{w}_{n,m,k} + \frac{\lambda_2[\Psi_{nk}\dot{u}_{n,m,k} + \Psi_{mk}\dot{v}_{n,m,k} + (\Phi_{mk}+\Phi_{nk})\dot{w}_{n,m,k}]}{2}
\end{aligned}\right\} \quad (2.2)$$

where $M$ is the mass of a block; $k_1$ and $\lambda_1$ are the spring stiffness and damper viscosity in the directions of the axes $x$, $y$, $z$; $k_2$ and $\lambda_2$ are the spring stiffness and damper viscosity in the diagonal directions. In the particular case, when $\lambda_1 = \lambda_2 = 0$, the lattice model (2.2) coincides with the special case of a model of Born & v. Kármán lattice [23], which does not take into account the shear stiffness.

In order to derive the equations of the motion at the boundary points, we consider that the stresses are equal to zero on the free surface. More precisely, using the notation

$$\left.\begin{aligned}
&\Lambda_k^- f_{n,m,0} = f_{n+1,m,-1} - f_{n,m,0}, \\
&\Phi_{nk}^- f_{n,m,0} = f_{n-1,m,-1} - 2f_{n,m,0} + f_{n+1,m,-1}, \quad \Phi_{mk}^- f_{n,m,0} = f_{n,m-1,-1} - 2f_{n,m,0} + f_{n,m+1,-1} \\
\text{and}\quad &\Psi_{nk}^- f_{n,m,0} = f_{n-1,m,-1} - f_{n+1,m,-1}, \quad \Psi_{mk}^- f_{n,m,0} = f_{n,m-1,-1} - f_{n,m+1,-1},
\end{aligned}\right\}$$

we write the equations of the motion of the mass with indices $n$, $m$, 0, located on the surface of the half-space in the following form:

$$\left.\begin{aligned}
M\ddot{u}_{n,m,0} &= k_1\Lambda_{nn}u_{n,m,0} + \frac{k_2[(\Phi_{nk}^-+\Phi_{nm}^-)u_{n,m,0} + \Psi_{nm}^-v_{n,m,0} + \Psi_{nk}^-w_{n,m,0}]}{2} \\
&\quad + \lambda_1\Lambda_{nn}\dot{u}_{n,m,0} + \frac{\lambda_2[(\Phi_{nk}^-+\Phi_{nm}^-)\dot{u}_{n,m,0} + \Psi_{nm}^-\dot{v}_{n,m,0} + \Psi_{nk}^-\dot{w}_{n,m,0}]}{2} \\
M\ddot{v}_{n,m,0} &= k_1\Lambda_{mm}v_{n,m,0} + \frac{k_2[\Psi_{nm}^-u_{n,m,0} + (\Phi_{mk}^-+\Phi_{nm}^-)v_{n,m,0} + \Psi_{mk}^-w_{n,m,0}]}{2} \\
&\quad + \lambda_1\Lambda_{mm}\dot{v}_{n,m,0} + \frac{\lambda_2[\Psi_{nm}^-\dot{u}_{n,m,0} + (\Phi_{mk}^-+\Phi_{nm}^-)\dot{v}_{n,m,0} + \Psi_{mk}^-\dot{w}_{n,m,0}]}{2} \\
\text{and}\quad M\ddot{w}_{n,m,0} &= k_1\Lambda_k^- w_{n,m,0} + \frac{k_2[\Psi_{nk}^-u_{n,m,0} + \Psi_{mk}^-v_{n,m,0} + (\Phi_{mk}^-+\Phi_{nk}^-)w_{n,m,0}]}{2} \\
&\quad + \lambda_1\Lambda_k^- \dot{w}_{n,m,0} + \frac{\lambda_2[\Psi_{nk}^-\dot{u}_{n,m,0} + \Psi_{mk}^-\dot{v}_{n,m,0} + (\Phi_{mk}^-+\Phi_{nk}^-)\dot{w}_{n,m,0}]}{2} + Q(t)\delta_{n0}\delta_{m0},
\end{aligned}\right\} \quad (2.3)$$

where $\delta_{0n}$ is the Kronecker delta, and $Q(t)$ is intensity of the vertical point load applied on the surface of the block half-space.

The parameters $M$, $k_1$, $k_2$, $\lambda_1$, $\lambda_2$ appearing in (2.2) and (2.3) have the same values for all points of the medium. Derivation of equations (2.2) and (2.3) can be found in the electronic supplementary material.

Since the planes $n=0$ and $m=0$ are the planes of symmetry of the wave process, the following relations are fulfilled ($k \leq 0$):

$$n=0: \quad u_{n,m,k} = -u_{-n,m,k}, \quad v_{n,m,k} = v_{-n,m,k}, \quad w_{n,m,k} = w_{-n,m,k}$$
$$\text{and} \quad m=0: \quad u_{n,m,k} = u_{n,-m,k}, \quad v_{n,m,k} = -v_{n,-m,k}, \quad w_{n,m,k} = w_{n,-m,k}. \tag{2.4}$$

The initial conditions for equations (2.2) and (2.3) are supposed to be zero.

## 3. Dispersion properties

Let $l$ be the length of a spring in the direction of the axis. In (2.2), replace each difference expression (2.1) by its differential approximation, e.g.

$$\Lambda_{nn} u_{n,m,k} = u_{n+1,m,k} - 2u_{n,m,k} + u_{n-1,m,k} \approx l^2 \frac{\partial^2 u}{\partial x^2} + \frac{l^4}{12}\frac{\partial^4 u}{\partial x^4} + \ldots$$

Assuming $l \to 0$, omit the summands of the orders higher than $l^2$. This transformation corresponds to the passage from the block medium to the homogeneous elastic medium. As a result we see that, for $l$ small enough, equations (2.2) turn into the equations of the orthotropic elasticity theory:

$$M\ddot{u} = l^2\left[(k_1+2k_2)\frac{\partial^2 u}{\partial x^2} + k_2\left(\frac{\partial^2 u}{\partial y^2}+\frac{\partial^2 u}{\partial z^2}\right) + 2k_2\left(\frac{\partial^2 v}{\partial x \partial y}+\frac{\partial^2 w}{\partial x \partial z}\right)\right],$$
$$M\ddot{v} = l^2\left[(k_1+2k_2)\frac{\partial^2 v}{\partial y^2} + k_2\left(\frac{\partial^2 v}{\partial x^2}+\frac{\partial^2 v}{\partial z^2}\right) + 2k_2\left(\frac{\partial^2 u}{\partial x \partial y}+\frac{\partial^2 w}{\partial y \partial z}\right)\right],$$
$$\text{and} \quad M\ddot{w} = l^2\left[(k_1+2k_2)\frac{\partial^2 w}{\partial z^2} + k_2\left(\frac{\partial^2 w}{\partial y^2}+\frac{\partial^2 w}{\partial z^2}\right) + 2k_2\left(\frac{\partial^2 u}{\partial x \partial z}+\frac{\partial^2 v}{\partial y \partial z}\right)\right]. \tag{3.1}$$

If $k_1 = k_2$, equations (3.1) describe motions of an isotropic elastic medium with Poisson's ratio $\sigma = 0.25$, which corresponds to the following velocities of longitudinal ($c_\text{p}$) and shear ($c_\text{s}$) waves:

$$c_\text{p} = l\sqrt{\frac{3k_1}{M}} \quad \text{and} \quad c_\text{s} = l\sqrt{\frac{k_1}{M}}. \tag{3.2}$$

Below we always assume $k_1 = k_2$ and $\lambda_1 = \lambda_2$. The Rayleigh equation for the elastic medium is as follows [28]:

$$\left(2-\frac{c^2}{c_s^2}\right)^2 - 4\sqrt{\left(1-\frac{c^2}{c_p^2}\right)\left(1-\frac{c^2}{c_s^2}\right)} = 0.$$

Using the above equation and (3.2), we calculate the velocity of the Rayleigh waves of an isotropic elastic medium with Poisson's ratio $\sigma = 0.25$:

$$c_R = l\sqrt{\frac{2k_1}{M}\left(1-\frac{1}{\sqrt{3}}\right)}. \tag{3.3}$$

In order to obtain the dispersion relations for the model described by equation (2.2), we apply to each of these equations the Laplace transform (of parameter $p$) with respect to time $t$ (denoted by the superscript L), and the discrete Fourier transforms (of parameters $q_x, q_y, q_z$) with respect to variables $n, m, k$ respectively (denoted by the superscripts $F_n, F_m, F_k$). Put by definition:

$$U = (u_{n,m,k})^{LF_nF_mF_k}, \quad V = (v_{n,m,k})^{LF_nF_mF_k}, \quad W = (w_{n,m,k})^{LF_nF_mF_k} \quad \text{and} \quad \tilde{M} = \frac{Mp^2}{k_1 + \lambda_1 p}.$$

The Laplace–Fourier transform of equation (2.2) is as follows:

$$\begin{pmatrix} a_{11} + \tilde{M} & a_{12} & a_{13} \\ a_{21} & a_{22} + \tilde{M} & a_{23} \\ a_{31} & a_{32} & a_{33} + \tilde{M} \end{pmatrix} \begin{pmatrix} U \\ V \\ W \end{pmatrix} = \begin{pmatrix} 0 \\ 0 \\ 0 \end{pmatrix},$$

where

$$a_{11} = 4Z + 4Y + 4X(3 - 2Z - 2Y), \quad a_{22} = 4Z + 4X + 4Y(3 - 2Z - 2X),$$
$$a_{33} = 4Y + 4X + 4Z(3 - 2Y - 2X), \quad a_{12}^2 = a_{21}^2 = 64X(1-X)Y(1-Y),$$
$$a_{13}^2 = a_{31}^2 = 64X(1-X)Z(1-Z), \quad a_{23}^2 = a_{32}^2 = 64Y(1-Y)Z(1-Z),$$
$$X = \sin^2\left(\frac{q_x l}{2}\right), \quad Y = \sin^2\left(\frac{q_y l}{2}\right), \quad Z = \sin^2\left(\frac{q_z l}{2}\right).$$

Substituting $p = i\omega$ to the dispersion operator $\Delta = \det(A + \tilde{M}J)$, where $A$ is a matrix whose entries $a_{ij}$ are defined above, $J$ is the identity matrix, we obtain the dispersion equation for determining the frequency $\omega$ versus the wavenumbers $q_x, q_y, q_z$:

$$\Delta = \det\left(A - \frac{M\omega^2 J}{k_1 + i\lambda_1\omega}\right) = 0. \tag{3.4}$$

The roots of the cubic equation (3.4) can be found by Cardano's formula.

Let us investigate the dispersion equation (3.4) in the special case $\lambda_1 = 0$. Note that in this case, equation (3.4) coincides with the dispersion equation, studied in [24] for a three-dimensional lattice.

Let $q_z = 0$ or $q_z = \pi/l$. Then $Z = 0$ or $Z = 1$, $a_{13} = a_{23} = 0$, and the determinant in (3.4) splits into two factors:

$$\Delta = \left(a_{33} - \frac{M\omega^2}{k_1}\right)\left[\left(a_{11} - \frac{M\omega^2}{k_1}\right)\left(a_{22} - \frac{M\omega^2}{k_1}\right) - a_{12}^2\right] = 0 . \qquad (3.5)$$

The dispersion equation (3.5) has the following roots:

$$\text{if } q_z = 0 \quad \text{then} \quad \begin{cases} \omega_1 = 2\omega_0 \sqrt{3X + 3Y - 4XY}, \\ \omega_{2,3} = 2\omega_0 \sqrt{X + Y} \end{cases} \qquad (3.6)$$

and

$$\text{if } q_z = \frac{\pi}{l} \quad \text{then} \quad \begin{cases} \omega_1 = 2\omega_0 \sqrt{3 - Y - X}, \\ \omega_{2,3} = 2\omega_0 \sqrt{1 + X + Y - 2XY \pm 2\sqrt{Y(1-Y)X(1-X)}}, \end{cases} \qquad (3.7)$$

where $\omega_0 = \sqrt{k_1/M}$.

Since $A$ is symmetric, all its eigenvalues are real. Numerical calculations show that all the principal minors of the matrix $A$ are nonnegative for $\lambda_1 = 0$. Therefore, all the roots of the dispersion equation (3.4) are also nonnegative, i.e. $M\omega^2/k_1 \geq 0$. Hence, the frequency $\omega$ takes only real values for $\lambda_1 = 0$.

The vectors of the phase and group velocities and their norms are as follows:

$$c_f = \frac{\omega}{q_x^2 + q_y^2 + q_z^2}(q_x, q_y, q_z) \quad \text{and} \quad \left|c_f\right| = \frac{\omega}{\sqrt{q_x^2 + q_y^2 + q_z^2}} \qquad (3.8)$$

and

$$c_g = \left(\frac{\partial \omega}{\partial q_x}, \frac{\partial \omega}{\partial q_y}, \frac{\partial \omega}{\partial q_z}\right) \quad \text{and} \quad \left|c_g\right| = \sqrt{\left(\frac{\partial \omega}{\partial q_x}\right)^2 + \left(\frac{\partial \omega}{\partial q_y}\right)^2 + \left(\frac{\partial \omega}{\partial q_z}\right)^2} . \qquad (3.9)$$

Denote by $c_{f,j}$ the vector of the phase velocity corresponding to the frequency $\omega_j$, where $j = 1, 2, 3$. Using formulae (3.6) and (3.8), we can calculate the values of $\left|c_{f,j}\right|$ for $q_z = 0$. First of all we are interested in the velocities of the long waves. Therefore, we give the values of $\left|c_{f,j}\right|$ under the assumption that $q_x, q_y \to 0$ and $q_z = 0$:

$$\left|c_{f,1}\right| = l\sqrt{\frac{3k_1}{M}} = c_p \quad \text{and} \quad \left|c_{f,2}\right| = \left|c_{f,3}\right| = l\sqrt{\frac{k_1}{M}} = c_s . \qquad (3.10)$$

Note that, as might be expected, formulae (3.2) and (3.10) coincide with each other.

For arbitrary values of $q_x, q_y, q_z$, the norms of the vectors of the phase and group velocities were calculated numerically using formulae (3.4), (3.8) and (3.9). These numerical calculations show that the norms of the vectors of the phase and group velocities of the long waves in the block medium coincide with each other. This fact implies that infinitely long waves propagate in the block medium without dispersion and form low-frequency pendulum waves.

Numerical calculations show that the vector of the group velocity is equal to zero at the following points:

$$(q_x l, q_y l, q_z l) = (0, 0, \pi), (0, \pi, 0), (\pi, 0, 0), (0, \pi, \pi), (\pi, 0, \pi), (\pi, \pi, 0), (\pi, \pi, \pi).$$

At these points, the frequencies $\omega_j$ ($j = 1, 2, 3$) are the eigenfrequencies of the short waves in the three-dimensional block medium. It follows from (3.6) and (3.7) that the eigenfrequencies are as follows:

$$\omega_1 = 2\sqrt{3}\omega_0, \quad \omega_{2,3} = 2\omega_0, \quad \text{if} \quad (q_x l, q_y l, q_z l) = (0, 0, \pi), (0, \pi, 0), (\pi, 0, 0),$$

$$\omega_{1,2,3} = 2\sqrt{2}\omega_0, \quad \text{if} \quad (q_x l, q_y l, q_z l) = (0, \pi, \pi), (\pi, 0, \pi), (\pi, \pi, 0),$$

$$\omega_{1,2,3} = 2\omega_0, \quad \text{if} \quad (q_x l, q_y l, q_z l) = (\pi, \pi, \pi).$$

It follows that there are only three different values of the eigenfrequency: $2\omega_0$, $2\sqrt{2}\omega_0$ and $2\sqrt{3}\omega_0$.

Figure 3 depicts graphs of the frequency versus wavenumbers $q_x, q_y$ for the following three values of $q_z$: $0, \pi/2, \pi$. Figure 4 shows graphs of the norm of the phase velocity versus wavenumbers $q_x, q_y$ for the same three values of $q_z$. Calculations, the results of which are shown on figures 3 and 4, are made for the lattice with the following values of the parameters: $l = M = 1$, $\lambda_1 = 0$, $k_1 = 0.4$. These figures show that, for $q_z = 0$, the dispersion equation has one double and one simple root. If $q_z = \pi/2, \pi$, for some values of $q_x, q_y$, there are three different simple roots, while, for some other values of $q_x, q_y$, there is one double root and one simple root. If $q_x = q_y = \pi$, there is one root of the multiplicity 3 for all values of $q_z$. From the dispersion equation (3.4), it follows that $\omega$ and $c_f$ are symmetric functions of the wavenumbers $q_x, q_y, q_z$. Therefore, the above conclusions remain correct under any permutation of the variables $q_x, q_y, q_z$. Figure 4 shows that long-wave perturbations have the maximal velocity in the block medium. Short waves have a smaller velocity, so they will be lag behind the long waves.

**4. Stability of the difference scheme**

We solve equations (2.2) and (2.3) with conditions (2.4) and zero initial conditions by a finite difference method using an explicit scheme. For the second derivatives with respect to time we use the second central difference approximation of the second order of accuracy:

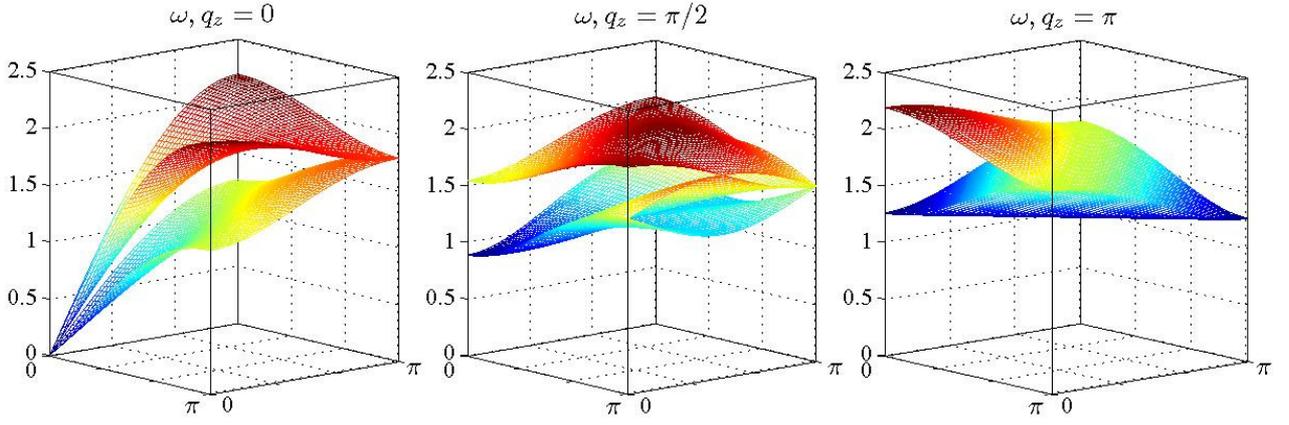

**Figure 3.** Dependencies of the frequency versus wavenumbers $q_x, q_y$ for the three values of $q_z$: $0, \pi/2, \pi$.

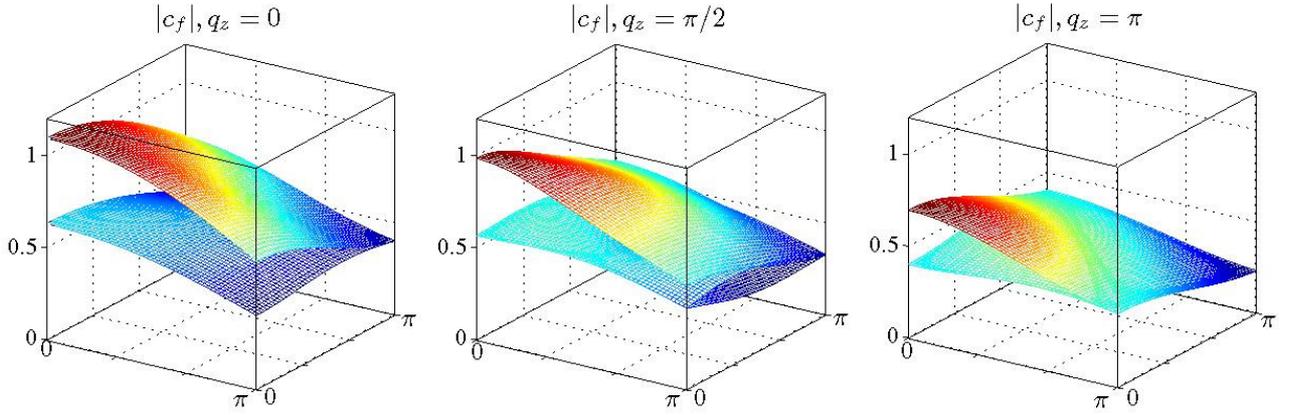

**Figure 4.** Dependence of the norm of the phase velocity versus wavenumbers $q_x, q_y$ for the three values of $q_z$: $0, \pi/2, \pi$.

$$\ddot{u}_{n,m,k} \approx \frac{u^{s+1}_{n,m,k} - 2u^s_{n,m,k} + u^{s-1}_{n,m,k}}{\tau^2}, \quad k \le 0, \quad s = 0,1,2,....$$

For the first derivatives with respect to time, we use the backward difference approximation of the first order of accuracy:

$$\dot{u}_{n,m,k} \approx \frac{u^s_{n,m,k} - u^{s-1}_{n,m,k}}{\tau}, \quad k \le 0, \quad s = 0,1,2,...,$$

where $\tau$ is the time step of the difference mesh, $u^s_{n,m,k}$ is the value of the displacement $u_{n,m,k}(t)$ at time $t = s\tau$, $s$ is the number of the time step in the finite difference scheme. We use the same approximations for the displacements $v_{n,m,k}(t)$ and $w_{n,m,k}(t)$. It can be shown that the stability condition of the difference equations, corresponding to equation (2.2), is as follows:

$$\tau \le -\frac{\lambda_1}{k_1} + \sqrt{\left(\frac{\lambda_1}{k_1}\right)^2 + \frac{M}{3k_1}}.$$

The stability condition of the difference equations, corresponding to equation (2.3), is as follows:

$$\tau \leq -\frac{\lambda_1}{k_1} + \sqrt{\left(\frac{\lambda_1}{k_1}\right)^2 + \frac{2M}{5k_1}}.$$

Thus, the stability condition of the difference equations of Lamb's problem for the block medium is as follows:

$$\tau \leq \min\left\{-\frac{\lambda_1}{k_1} + \sqrt{\left(\frac{\lambda_1}{k_1}\right)^2 + \frac{M}{3k_1}}, -\frac{\lambda_1}{k_1} + \sqrt{\left(\frac{\lambda_1}{k_1}\right)^2 + \frac{2M}{5k_1}}\right\} = -\frac{\lambda_1}{k_1} + \sqrt{\left(\frac{\lambda_1}{k_1}\right)^2 + \frac{M}{3k_1}}.$$

Numerical calculations confirm the validity of this stability condition.

## 5. Results of numerical experiments: the case of a step load

In this section, we present the results of numerical calculations of disturbances in Lamb's problem for a lattice under the action of the vertical step load $Q(t) = Q_0 H(t)$ applied at the point with coordinates $(0,0,0)$. Here $H(t)$ is the Heaviside step-function and $Q_0$ is the amplitude of the load. All calculations, the results of which are presented in this section, are done for the case of $Q_0 = l = M = 1$, $k_1 = 0.4$, $\tau = 0.5$. Denote by $w = w_{n,m,0}$, $u_r = (nu_{n,m,0} + mv_{n,m,0})/\sqrt{n^2 + m^2}$ the vertical and radial displacements on the surface of the half-space in cylindrical coordinate system $r, \theta, z$ with the origin at the location of the load.

In [41], an analytical solution is given to static spatial Lamb's problem for an isotropic elastic medium which is known as Boussinesq's problem. In our notation, the solution of the Boussinesq's problem, given in [41], is as follows on the surface $k = 0$:

$$u_r(n,m,0) = \frac{Q_0(1-2\sigma)(1+\sigma)}{2\pi E \sqrt{n^2 + m^2}} \quad \text{and} \quad w(n,m,0) = \frac{Q_0(1-\sigma^2)}{\pi E \sqrt{n^2 + m^2}}, \tag{5.1}$$

where $E$ is the Young modulus and $\sigma$ is Poisson's ratio.

For $l \to 0$, equation (2.2), describing the motions of the block medium, turn to equation (3.1), describing the motions of an isotropic elastic medium with Poisson's ratio $\sigma = 0.25$. For this reason, we put $\sigma = 0.25$ and $Q_0/E = 1$ in (5.1) in order to compare numerical solutions for the block medium with solution (5.1) for the elastic medium.

Figure 5 shows the time dependence of the radial $u_r$ and vertical $w$ displacements and their velocities $\dot{u}_r$ and $\dot{w}$ on the surface of the half-space at the point with the coordinates $(40,0,0)$, for different values of viscosity $\lambda_1$. On the plots of the displacements $u_r$ and $w$, shown in figure 5 (as well as in figure 7), the horizontal dashed lines correspond to the static solution (5.1). The vertical dashed lines in figure 5 (as well as in figures 7–11) correspond to the arrival time of the quasi-fronts of the longitudinal, shear, and Rayleigh waves at the point with the coordinates $(n,0,0)$: $t_p = n/c_p$, $t_s = n/c_s$, $t_R = n/c_R$, where $c_p, c_s, c_R$ are defined in (3.2) and (3.3).

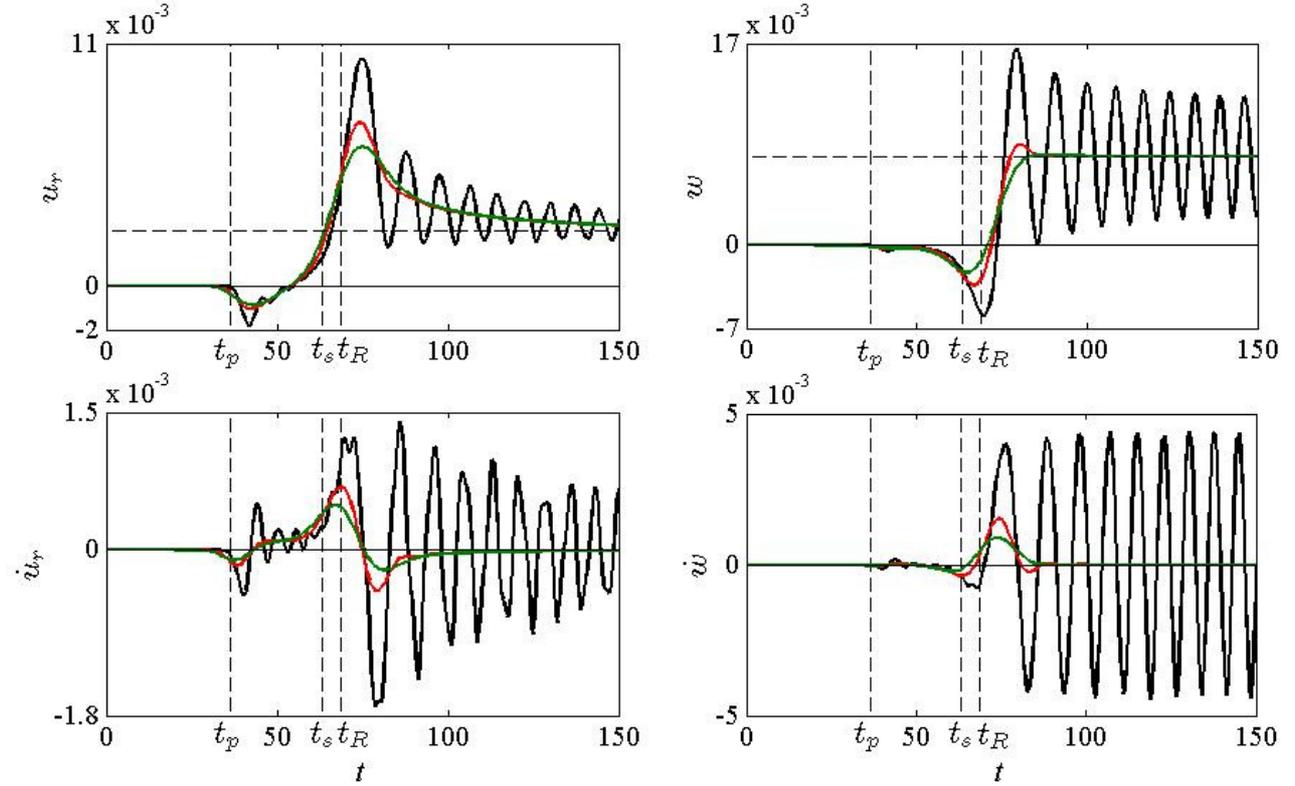

**Figure 5.** Radial $u_r$ and vertical $w$ displacements and their velocities $\dot{u}_r$, $\dot{w}$ on the surface of the half-space calculated at the point $n = 40, m = 0$. Black curves correspond to $\lambda_1 = 0$, red curves correspond to $\lambda_1 = 0.1$, and green curves correspond to $\lambda_1 = 0.2$.

As can be seen from figure 5, the displacements $u_r$ and $w$ behind the front of the Rayleigh wave oscillate, with frequency $\omega \approx \sqrt{5k_1/(2M)} = 1$, near the static solutions (5.1) for the elastic medium, while the velocities $\dot{u}_r$ and $\dot{w}$ of the displacements oscillate near zero. Observe that the maximal amplitude of the velocities $\dot{u}_r$ and $\dot{w}$ of the displacements occur in the vicinity of the front of the low-frequency Rayleigh wave. Figure 5 shows than an increase in viscosity reduces the amplitude of high-frequency oscillations and then to their extinction. Note that the high-frequency oscillations of the perturbations in the vicinity of the fronts of the longitudinal and Rayleigh waves, which can be seen in figure 5, are caused by dispersion inherent to the model block medium (see §3), not by numerical dispersion. This follows from the fact that (i) numerical calculations, carried out for smaller values of the time step ($\tau = 0.25$), result in the same plots of the solutions as presented in figure 5 and (ii) Alessandrini [20] has shown that similar high-frequency oscillations are present in the solution to Lamb's problem for a three-dimensional discrete medium (Alessandrini studied a discrete medium and used a method that completely different from those in our paper), and (iii) in the two-dimensional case, the author has shown [16,18] that this is dispersion of the discrete model, not numerical dispersion. Let us explain the last argument in more detail. The studies [16,18] are devoted to solving two problems for the two-dimensional discrete periodic media. In [16,18], both numerical and analytical solutions are obtained for a step point load and these solutions are shown to coincide with each other both qualitatively and quantitatively, while the high-frequency oscillations are caused by the short-wave perturbations with the wavenumber $ql = \pi$.

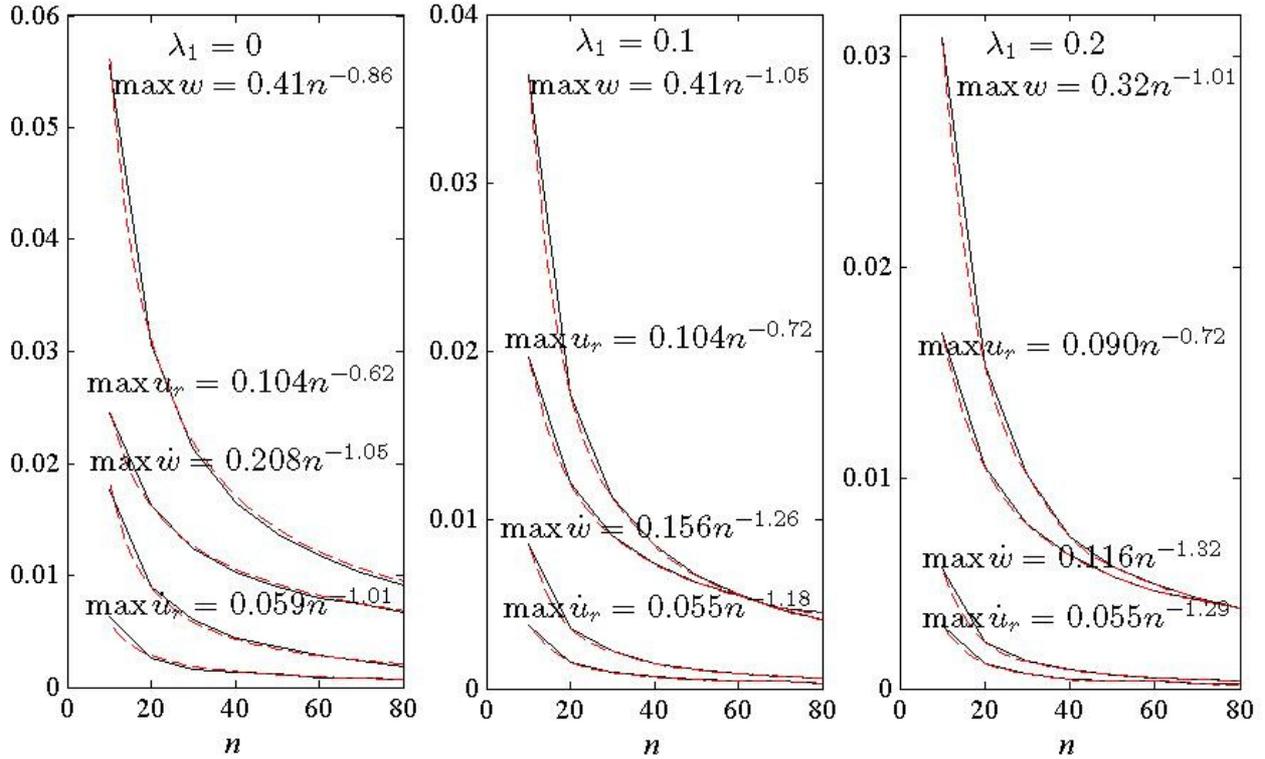

**Figure 6.** Dependence of the maximal amplitude of the $u_r$, $w$, $\dot{u}_r$ and $\dot{w}$ versus the coordinate $n$ for different values of viscosity $\lambda_1$. Black curves depict the results of the numerical calculations. Red dashed curves show their approximations of the form $f = An^\alpha$ (concrete formulae are written next to the curves).

More generally we can say that numerical dispersion appearing in the transition from the equations of the theory of elasticity to their difference analogues is of the same nature as dispersion inherent to the discrete periodic media. However, we have to reduce parasitic effects of the discretization, while we have to take into consideration dispersion inherent to the block medium and to study its impact on the wave process.

Solid curves in figure 6 depicts the plots of the maximal amplitudes of the perturbations $u_r$, $w$, $\dot{u}_r$, and $\dot{w}$ versus the coordinate $n$ calculated for $m = k = 0$ and different values of the viscosity parameter: $\lambda_1 = 0$, $\lambda_1 = 0.1$, and $\lambda_1 = 0.2$. We approximate each solid curve by a plot of a power function $f = An^\alpha$, where the constants $A$ and $\alpha$ are such that the values of $f$ at the points $n = 10, 20, 30, ..., 80$ are as close as possible to the values of the corresponding function $u_r$, $w$, $\dot{u}_r$, or $\dot{w}$ at the same points. The dashed curves in figure 6 show the plots of these functions $f = An^\alpha$, the formulae defining $f$ are written next to the curve. Figure 6 shows that the maximal amplitude of the velocities $\dot{u}_r$ and $\dot{w}$ of the displacements as a function of $n$, tends to zero as $n$ is increasing. Moreover, the rate of the descent increases with the increasing of the viscosity of the interlayers.

Thus, figures 5 and 6 show that the dissipative properties of the interlayers result in additional damping of the low-frequency pendulum waves and in more rapid decay of the high-frequency waves, which is more consistent with real seismograms.

In table 1, we present the power-law dependences of the damping of the amplitudes of the disturbances versus coordinate $n$, corresponding to $\lambda_1 = 0$ (figure 6). In the same table, for comparison, we give similar results for two-dimensional Lamb's problem for a block medium [18]. As the table shows, the degree of the descent of the maximum amplitudes of the perturbations in the block medium in three-dimensional Lamb's problem is significantly higher than in two-dimensional Lamb's problem.

**Table 1.** Degree of the descent of the maximum amplitudes of the perturbations versus $n$ in the block medium (without taking into account the viscosity of the interlayers) in two-dimensional and three-dimensional Lamb's problems.

|  | max $w$ | max $u_r$ | max $\dot{w}$ | max $\dot{u}_r$ |
|---|---|---|---|---|
| two-dimensional | $\ln n$ | $n^0$ | $n^{-1/3}$ | $n^{-1/3}$ |
| three-dimensional | $n^{-5/6}$ | $n^{-2/3}$ | $n^{-1}$ | $n^{-1}$ |

Figure 7 shows plots of the quantities $u_r$, $w$, $\dot{u}_r$ and $\dot{w}$ versus time which are obtained, first, numerically for the bock medium at the point with the coordinates $n = 30, m = 0$ and, second, from an analytical solution, given in [39], for the homogeneous elastic medium with Poisson's ratio $\sigma = 0.25$ at a point with the coordinate $r = 30$. All the quantities, whose plots are shown in figure 7, are calculated on the surface of a half-space for the case of the vertical step load.

In the electronic supplementary material to [39], a program is given for calculating the vertical and radial displacements on the surface of an elastic half-space. Using that program, we calculate the vertical $w$ and radial $u_r$ displacements on the surface of the elastic half-space. Then we numerically differentiate the displacements with respect to time and find the radial $\dot{u}_r$ and vertical $\dot{w}$ velocities of the displacements at the same point. The plots of these quantities are shown in figure 7. We see that the solution for the block medium differs significantly from the solution for the elastic medium. Nevertheless, these solutions are in qualitative agreement, namely, the solution for the block medium oscillates near the solution for the elastic medium.

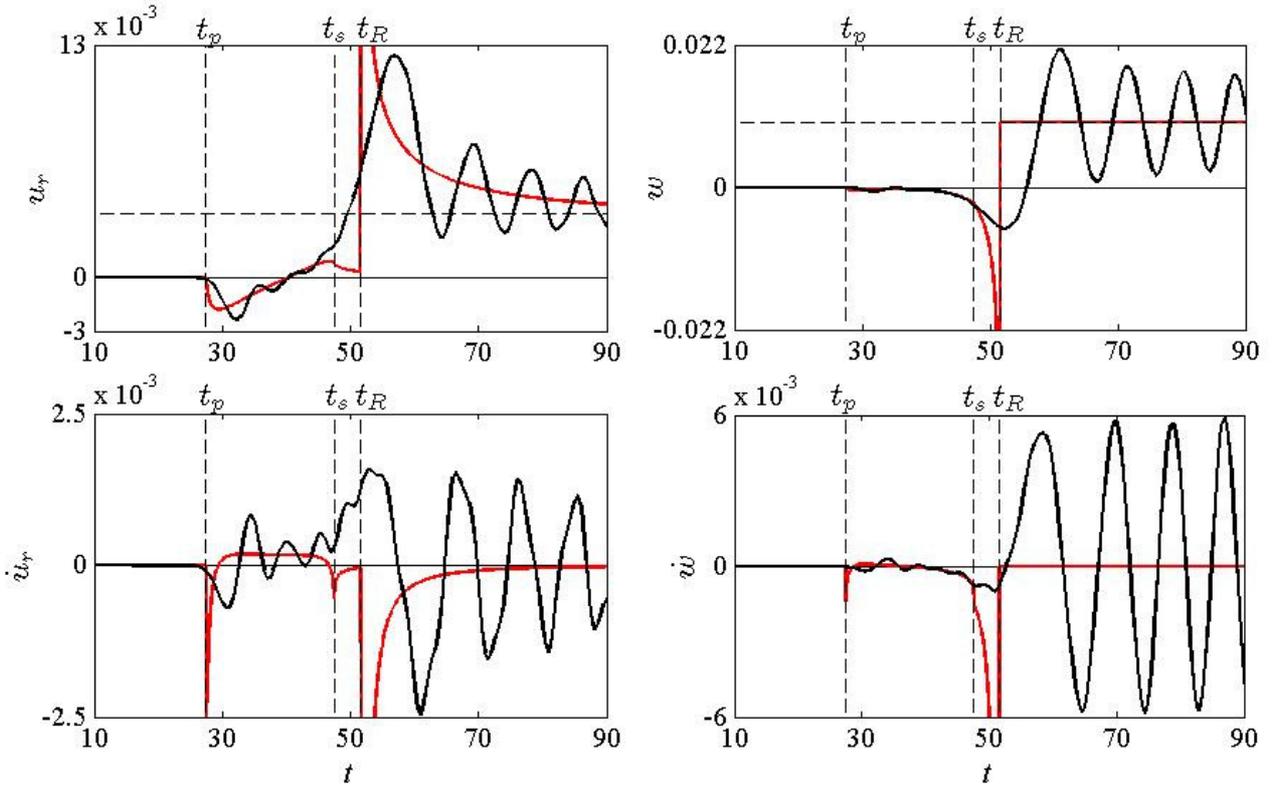

**Figure** 7. Radial $u_r$ and vertical $w$ displacements and their velocities $\dot{u}_r$ and $\dot{w}$, calculated at a point on the surface of the half-space for the block medium (black curves) and elastic medium (red curves).

## 6. Results of numerical experiments: the case of an impulse load

In this section, we present the results of numerical calculations of disturbances in Lamb's problem for a block medium under the action of the vertical impulse load given by the formula

$$Q(t) = Q_0 H(t)\sin(\omega t) H(t_0 - t) \tag{6.1}$$

and applied at the point with coordinates $(0,0,0)$. Here $t_0$ is duration of the impulse and $\omega = \pi/t_0$.

Figures 8–11 show the plots of the radial $u_r$ and vertical $w$ displacements and their velocities $\dot{u}_r$ and $\dot{w}$ versus time, calculated on the surface of the half-space for the block medium at the point $n = 40$, $m = 0$ and for the elastic medium at a point with $r = 40$. All calculations are done for two values of $\omega$, namely, $\omega = 0.2$ and $\omega = 0.1$, and the following values of the other parameters: $Q_0 = l = M = 1$, $k_1 = 0.4$, $\lambda_1 = 0$, $\sigma = 0.25$ and $\tau = 0.5$.

We use Duhamel's integral

$$f(t) = \int_0^t g(\tau)\frac{\partial Q(t-\tau)}{\partial t} d\tau, \tag{6.2}$$

where $g(\tau)$ and $f(t)$ are the response functions owing to the step load and impulse load (6.1), respectively, and, as in §5, the program, given in the electronic supplementary material to [39], for calculating the vertical and radial displacements on the surface of an elastic half-space. We use the trapezoidal rule in order to calculate the integral in (6.2).

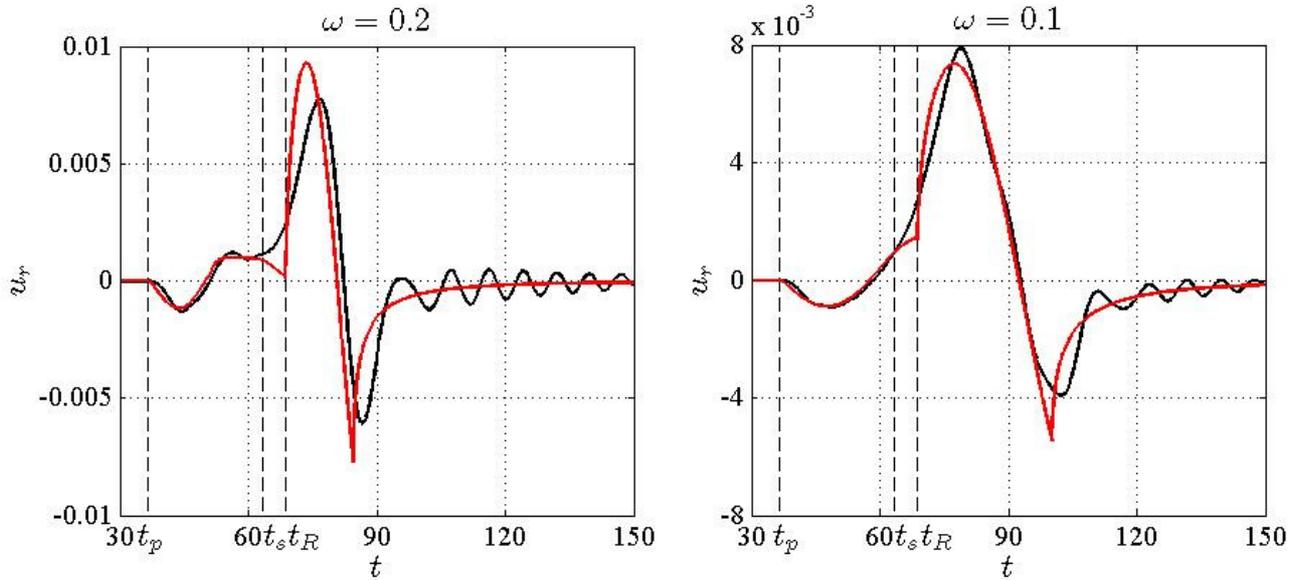

**Figure 8.** Radial displacement $u_r$ on the surface of the half-space for the block medium calculated at the point $n = 40, m = 0$ (black curves) and for the elastic medium calculated at a point with $r = 40$ (red curves).

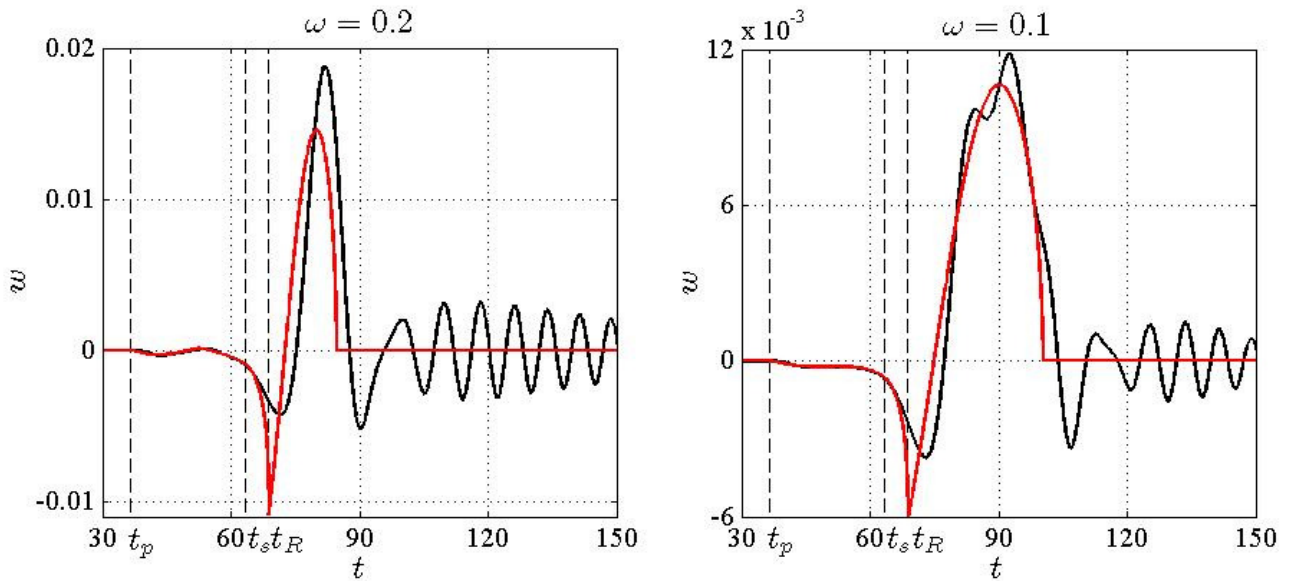

**Figure 9.** Vertical displacement $w$ on the surface of the half-space for the block medium calculated at the point $n = 40, m = 0$ (black curves) and for the elastic medium calculated at a point with $r = 40$ (red curves).

Figures 8–11 show that the qualitative and quantitative agreement between the results obtained for the block and elastic media are improved as the duration of the impulse (6.1) increases. The best agreement takes place for the displacements $u_r$ and $w$ (figures 8 and 9).

The results of numerical calculations for loads which have time dependencies different from those considered in §5 and 6 can be found in the electronic supplementary material.

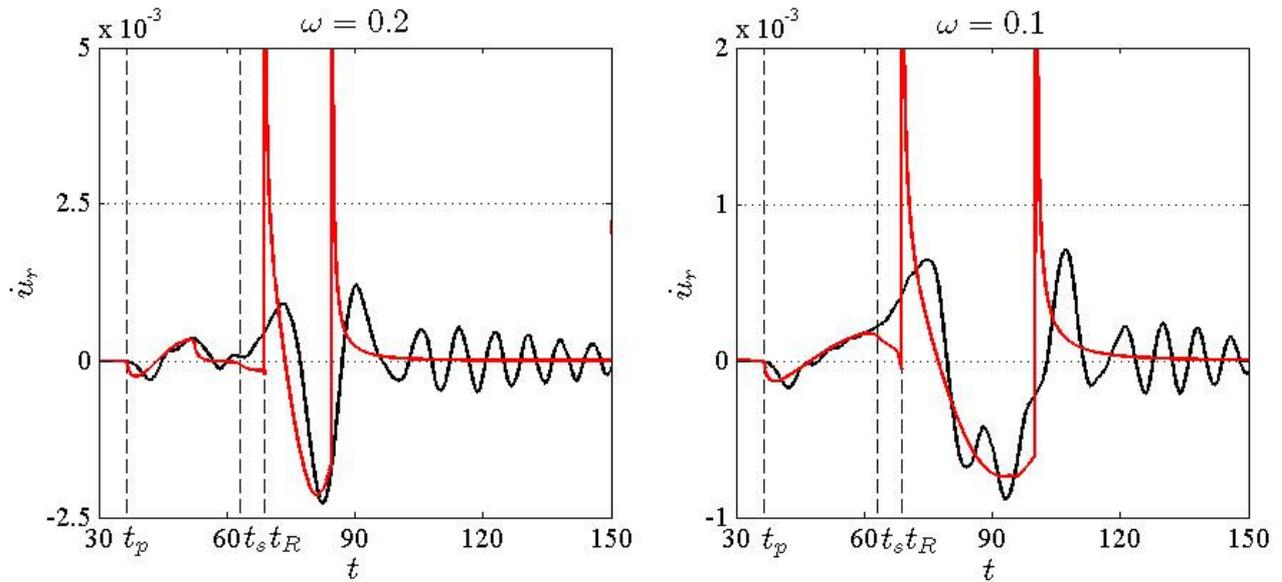

**Figure 10.** Radial velocity $\dot{u}_r$ of the displacement on the surface of the half-space for the block medium calculated at the point $n = 40, m = 0$ (black curves) and for the elastic medium calculated at a point with $r = 40$ (red curves).

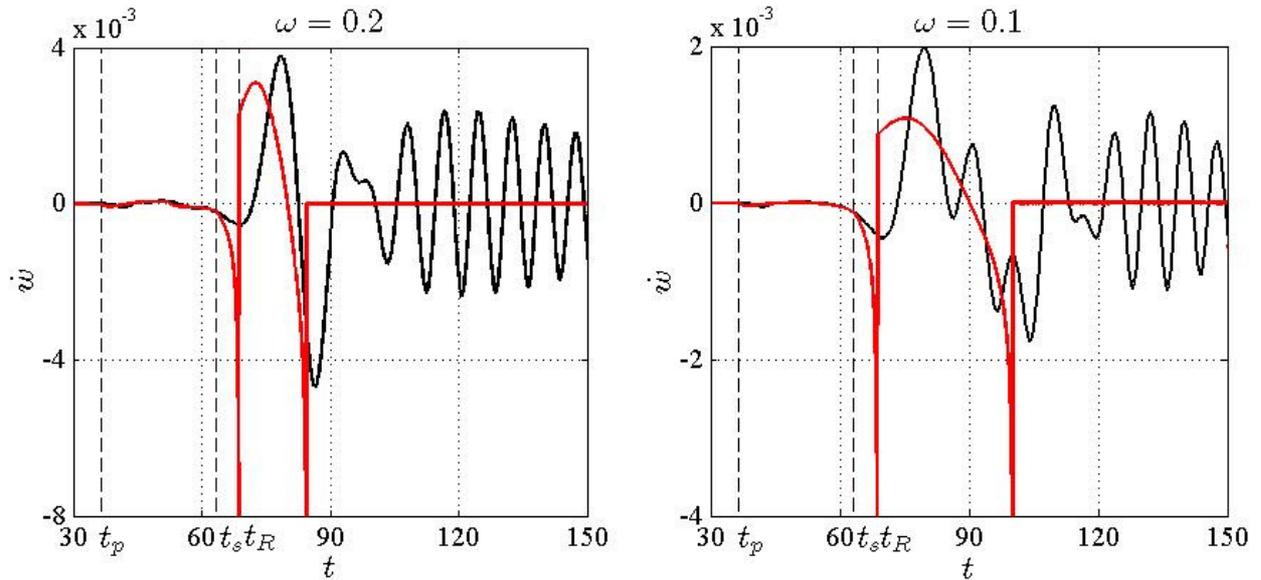

**Figure 11.** Vertical velocity $\dot{w}$ of the displacement on the surface of the half-space for the block medium calculated at the point $n = 40, m = 0$ (thick curves) and for the elastic medium calculated at a point with $r = 40$ (thin curves).

## 7. Conclusion

In this paper, we proposed a three-dimensional mathematical model of block rock masses. This model is based on the idea that the dynamic behaviour of a block medium can be roughly described as the motion of rigid blocks due to compressibility of interlayers between them, and that the deformation of the interlayers can be approximately described by Kelvin–Voigt's model. Using this model, we solved Lamb's problem numerically, specifically, we studied the propagation of seismic waves in a block medium under the effect of a vertical transient point load applied at the surface of the half-space. For Lamb's problem, we have shown that the main contribution to the wave process on the surface of the block medium is made by low-frequency waves in the vicinity of the front of the Rayleigh wave and that high-frequency waves are observed behind the front of the Rayleigh wave.

It is shown that the presence of the block structure in the medium leads to the following changes in its behaviour in comparison with what the model predicts of a homogeneous elastic medium, whose mechanical properties are obtained by averaging the mechanical properties of the block medium:

- In the block medium, waves propagate with dispersion, which is absent in the homogeneous elastic medium.
- On the surface of the block medium, low-frequency longitudinal and Rayleigh waves propagate with velocities which are much smaller than the corresponding velocities in the blocks.
- The velocity of propagation of low-frequency waves and the degree of attenuation is determined by the weight of the blocks, their dimensions and properties of interlayers.
- The dissipative properties of the interlayers result in additional damping of the low-frequency waves on the surface of the half-space in the vicinity of the front of the Rayleigh wave and in more rapid decay of the high-frequency waves behind the front of the Rayleigh wave.
- The difference in the behaviour of the block medium from the behaviour of the homogeneous elastic medium is especially considerable for the case of a short impulse. When the impulse duration is increased, the difference in the behaviour of the block medium from the behaviour of the homogeneous elastic medium becomes less noticeable.

The results obtained in this article show that in the study of seismic waves, it is necessary to take into account the block structure of the rocks and rheological properties of the interlayers.

**Competing interests statement.** I declare I have no competing interests.
**Funding statement.** This work was not supported by a grant.

# Electronic supplementary material to the article

## 'Seismic waves in a three-dimensional block medium' by N.I. Aleksandrova

(*N.A. Chinakal Institute of Mining, Siberian Branch, Russian Academy of Sciences,*
*Krasnyi pr. 54, Novosibirsk 630091, Russia,* e-mail: *nialex@misd.ru*)

## 1. Derivation of equations of the motion of the three-dimensional lattice of masses

In this section, we derive the equations of the motion of the three-dimensional lattice of masses in Lamb's problem, whose model is shown in figures 1 and 2.

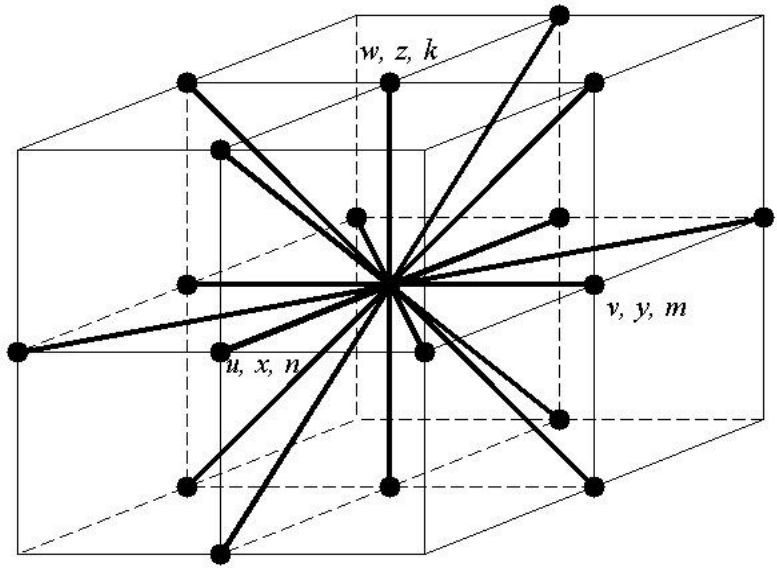

**Figure 1.** Scheme of connections of the masses by springs and dampers in the three-dimensional model of a block medium.

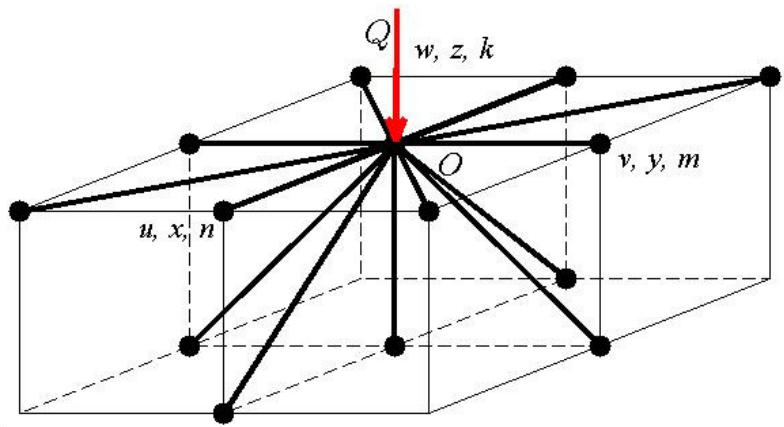

**Figure 2.** Scheme of connections of the masses by springs and dampers on the surface on the half-space in the three-dimensional model of a block medium and the point load.

Within the frames of Kelvin–Voigt's model [41], the forces with which the neighbouring masses act on the mass with the coordinates ($n, m, k$), are determined by the following formulas:

$$F_u^+ = k_1(u_{n+1,m,k} - u_{n,m,k}) + \lambda_1(\dot{u}_{n+1,m,k} - \dot{u}_{n,m,k}), \tag{1}$$

$$F_u^- = k_1(u_{n-1,m,k} - u_{n,m,k}) + \lambda_1(\dot{u}_{n-1,m,k} - \dot{u}_{n,m,k}),$$

$$F_v^+ = k_1(v_{n,m+1,k} - v_{n,m,k}) + \lambda_1(\dot{v}_{n,m+1,k} - \dot{v}_{n,m,k}),$$

$$F_v^- = k_1(v_{n,m-1,k} - v_{n,m,k}) + \lambda_1(\dot{v}_{n,m-1,k} - \dot{v}_{n,m,k}),$$

$$F_w^+ = k_1(w_{n,m,k+1} - w_{n,m,k}) + \lambda_1(\dot{w}_{n,m,k+1} - \dot{w}_{n,m,k}),$$

$$F_w^- = k_1(w_{n,m,k-1} - w_{n,m,k}) + \lambda_1(\dot{w}_{n,m,k-1} - \dot{w}_{n,m,k}),$$

$$Q_u^{++0} = k_2(u_{n+1,m+1,k} - u_{n,m,k}) + \lambda_2(\dot{u}_{n+1,m+1,k} - \dot{u}_{n,m,k}),$$

$$Q_u^{--0} = k_2(u_{n-1,m-1,k} - u_{n,m,k}) + \lambda_2(\dot{u}_{n-1,m-1,k} - \dot{u}_{n,m,k}),$$

$$Q_u^{+-0} = k_2(u_{n+1,m-1,k} - u_{n,m,k}) + \lambda_2(\dot{u}_{n+1,m-1,k} - \dot{u}_{n,m,k}),$$

$$Q_u^{-+0} = k_2(u_{n-1,m+1,k} - u_{n,m,k}) + \lambda_2(\dot{u}_{n-1,m+1,k} - \dot{u}_{n,m,k}),$$

$$Q_u^{+0+} = k_2(u_{n+1,m,k+1} - u_{n,m,k}) + \lambda_2(\dot{u}_{n+1,m,k+1} - \dot{u}_{n,m,k}),$$

$$Q_u^{-0-} = k_2(u_{n-1,m,k-1} - u_{n,m,k}) + \lambda_2(\dot{u}_{n-1,m,k-1} - \dot{u}_{n,m,k}),$$

$$Q_u^{+0-} = k_2(u_{n+1,m,k-1} - u_{n,m,k}) + \lambda_2(\dot{u}_{n+1,m,k-1} - \dot{u}_{n,m,k}),$$

$$Q_u^{-0+} = k_2(u_{n-1,m,k+1} - u_{n,m,k}) + \lambda_2(\dot{u}_{n-1,m,k+1} - \dot{u}_{n,m,k}),$$

$$Q_u^{0++} = k_2(u_{n,m+1,k+1} - u_{n,m,k}) + \lambda_2(\dot{u}_{n,m+1,k+1} - \dot{u}_{n,m,k}),$$

$$Q_u^{0--} = k_2(u_{n,m-1,k-1} - u_{n,m,k}) + \lambda_2(\dot{u}_{n,m-1,k-1} - \dot{u}_{n,m,k}),$$

$$Q_u^{0+-} = k_2(u_{n,m+1,k-1} - u_{n,m,k}) + \lambda_2(\dot{u}_{n,m+1,k-1} - \dot{u}_{n,m,k}),$$

$$Q_u^{0-+} = k_2(u_{n,m-1,k+1} - u_{n,m,k}) + \lambda_2(\dot{u}_{n,m-1,k+1} - \dot{u}_{n,m,k}),$$

$$Q_v^{++0} = k_2(v_{n+1,m+1,k} - v_{n,m,k}) + \lambda_2(\dot{v}_{n+1,m+1,k} - \dot{v}_{n,m,k}),$$

$$Q_v^{--0} = k_2(v_{n-1,m-1,k} - v_{n,m,k}) + \lambda_2(\dot{v}_{n-1,m-1,k} - \dot{v}_{n,m,k}),$$

$$Q_v^{+-0} = k_2(v_{n+1,m-1,k} - v_{n,m,k}) + \lambda_2(\dot{v}_{n+1,m-1,k} - \dot{v}_{n,m,k}),$$

$$Q_v^{-+0} = k_2(v_{n-1,m+1,k} - v_{n,m,k}) + \lambda_2(\dot{v}_{n-1,m+1,k} - \dot{v}_{n,m,k}),$$

$$Q_v^{+0+} = k_2(v_{n+1,m,k+1} - v_{n,m,k}) + \lambda_2(\dot{v}_{n+1,m,k+1} - \dot{v}_{n,m,k}),$$

$$Q_v^{-0-} = k_2(v_{n-1,m,k-1} - v_{n,m,k}) + \lambda_2(\dot{v}_{n-1,m,k-1} - \dot{v}_{n,m,k}),$$

$$Q_v^{+0-} = k_2(v_{n+1,m,k-1} - v_{n,m,k}) + \lambda_2(\dot{v}_{n+1,m,k-1} - \dot{v}_{n,m,k}),$$

$$Q_v^{+0-} = k_2(v_{n+1,m,k-1} - v_{n,m,k}) + \lambda_2(\dot{v}_{n+1,m,k-1} - \dot{v}_{n,m,k}),$$

$$Q_v^{0-+} = k_2(v_{n,m-1,k+1} - v_{n,m,k}) + \lambda_2(\dot{v}_{n,m-1,k+1} - \dot{v}_{n,m,k}),$$

$$Q_w^{++0} = k_2(w_{n+1,m+1,k} - w_{n,m,k}) + \lambda_2(\dot{w}_{n+1,m+1,k} - \dot{w}_{n,m,k}),$$

$$Q_w^{--0} = k_2(w_{n-1,m-1,k} - w_{n,m,k}) + \lambda_2(\dot{w}_{n-1,m-1,k} - \dot{w}_{n,m,k}),$$

$$Q_w^{+-0} = k_2(w_{n+1,m-1,k} - w_{n,m,k}) + \lambda_2(\dot{w}_{n+1,m-1,k} - \dot{w}_{n,m,k}),$$

$$Q_w^{-+0} = k_2(w_{n-1,m+1,k} - w_{n,m,k}) + \lambda_2(\dot{w}_{n-1,m+1,k} - \dot{w}_{n,m,k}),$$

$$Q_w^{+0+} = k_2(w_{n+1,m,k+1} - w_{n,m,k}) + \lambda_2(\dot{w}_{n+1,m,k+1} - \dot{w}_{n,m,k}),$$

$$Q_w^{-0-} = k_2(w_{n-1,m,k-1} - w_{n,m,k}) + \lambda_2(\dot{w}_{n-1,m,k-1} - \dot{w}_{n,m,k}),$$

$$Q_w^{+0-} = k_2(w_{n+1,m,k-1} - w_{n,m,k}) + \lambda_2(\dot{w}_{n+1,m,k-1} - \dot{w}_{n,m,k}),$$

$$Q_w^{-0+} = k_2(w_{n-1,m,k+1} - w_{n,m,k}) + \lambda_2(\dot{w}_{n-1,m,k+1} - \dot{w}_{n,m,k}),$$

$$Q_w^{0++} = k_2(w_{n,m+1,k+1} - w_{n,m,k}) + \lambda_2(\dot{w}_{n,m+1,k+1} - \dot{w}_{n,m,k}),$$

$$Q_w^{0--} = k_2(w_{n,m-1,k-1} - w_{n,m,k}) + \lambda_2(\dot{w}_{n,m-1,k-1} - \dot{w}_{n,m,k}),$$

$$Q_w^{0+-} = k_2(w_{n,m+1,k-1} - w_{n,m,k}) + \lambda_2(\dot{w}_{n,m+1,k-1} - \dot{w}_{n,m,k}),$$

$$Q_w^{0-+} = k_2(w_{n,m-1,k+1} - w_{n,m,k}) + \lambda_2(\dot{w}_{n,m-1,k+1} - \dot{w}_{n,m,k}).$$

The equations of the motion of the mass with the coordinates ($n$, $m$, $k$) are as follows:

$$M\ddot{u}_{n,m,k} = F_u^+ + F_u^- + \frac{Q_u^{+0+} + Q_u^{-0-} + Q_u^{+0-} + Q_u^{-0+}}{2} + \frac{Q_u^{++0} + Q_u^{--0} + Q_u^{+-0} + Q_u^{-+0}}{2} \\ + \frac{Q_v^{++0} + Q_v^{--0} - Q_v^{+-0} - Q_v^{-+0}}{2} + \frac{Q_w^{+0+} + Q_w^{-0-} - Q_w^{+0-} - Q_w^{-0+}}{2}, \quad (2)$$

$$M\ddot{v}_{n,m,k} = F_v^+ + F_v^- + \frac{Q_v^{0++} + Q_v^{0--} + Q_v^{0+-} + Q_v^{0-+}}{2} + \frac{Q_w^{0++} + Q_w^{0--} - Q_w^{0+-} - Q_w^{0-+}}{2} \\ + \frac{Q_u^{++0} + Q_u^{--0} - Q_u^{+-0} - Q_u^{-+0}}{2} + \frac{Q_v^{++0} + Q_v^{--0} + Q_v^{+-0} + Q_v^{-+0}}{2},$$

$$M\ddot{w}_{n,m,k} = F_w^+ + F_w^- + \frac{Q_v^{0++} + Q_v^{0--} - Q_v^{0+-} - Q_v^{0-+}}{2} + \frac{Q_w^{0++} + Q_w^{0--} + Q_w^{0+-} + Q_w^{0-+}}{2} \\ + \frac{Q_u^{+0+} + Q_u^{-0-} - Q_u^{+0-} - Q_u^{-0+}}{2} + \frac{Q_w^{+0+} + Q_w^{-0-} + Q_w^{+0-} + Q_w^{-0+}}{2}.$$

On the free surface $k = 0$ the following forces are equal to zero:

$$Q_u^{+0+} = 0, \quad Q_u^{-0+} = 0, \quad Q_u^{0++} = 0, \quad Q_u^{0-+} = 0, \quad (3)$$

$$Q_v^{+0+} = 0, \quad Q_v^{-0+} = 0, \quad Q_v^{0++} = 0, \quad Q_v^{0-+} = 0,$$

$$F_w^+ = 0, \quad Q_w^{+0+} = 0, \quad Q_w^{-0+} = 0, \quad Q_w^{0++} = 0, \quad Q_w^{0-+} = 0.$$

Using (3), we get that the equations of the motion of the mass with the coordinates (*n, m*) on the surface $k = 0$ are as follows:

$$M\ddot{u}_{n,m,0} = F_u^+ + F_u^- + \frac{Q_u^{-0-} + Q_u^{+0-}}{2} + \frac{Q_u^{++0} + Q_u^{--0} + Q_u^{+-0} + Q_u^{-+0}}{2}$$
$$+ \frac{Q_v^{++0} + Q_v^{--0} - Q_v^{+-0} - Q_v^{-+0}}{2} + \frac{Q_w^{-0-} - Q_w^{+0-}}{2}, \quad (4)$$

$$M\ddot{v}_{n,m,0} = F_v^+ + F_v^- + \frac{Q_v^{0--} + Q_v^{0+-}}{2} + \frac{Q_w^{0--} - Q_w^{0+-}}{2}$$
$$+ \frac{Q_u^{++0} + Q_u^{--0} - Q_u^{+-0} - Q_u^{-+0}}{2} + \frac{Q_v^{++0} + Q_v^{--0} + Q_v^{+-0} + Q_v^{-+0}}{2},$$

$$M\ddot{w}_{n,m,0} = F_w^- + \frac{Q_v^{0--} - Q_v^{0+-}}{2} + \frac{Q_w^{0--} + Q_w^{0+-}}{2} + \frac{Q_u^{-0-} - Q_u^{+0-}}{2} + \frac{Q_w^{-0-} + Q_w^{+0-}}{2} + Q(t)\delta_{n0}\delta_{m0}.$$

Substituting (1) into (2), we obtain the equations of the motion of a three-dimensional block medium with viscoelastic interlayers:

$$\begin{aligned}
M\ddot{u}_{n,m,k} &= k_1(u_{n+1,m,k} - 2u_{n,m,k} + u_{n-1,m,k}) \\
&+ k_2(u_{n+1,m,k+1} + u_{n-1,m,k-1} + u_{n+1,m,k-1} + u_{n-1,m,k+1} - 4u_{n,m,k})/2 \\
&+ k_2(u_{n+1,m+1,k} + u_{n-1,m-1,k} + u_{n+1,m-1,k} + u_{n-1,m+1,k} - 4u_{n,m,k})/2 \\
&+ k_2(v_{n+1,m+1,k} + v_{n-1,m-1,k} - v_{n+1,m-1,k} - v_{n-1,m+1,k})/2 \\
&+ k_2(w_{n+1,m,k+1} + w_{n-1,m,k-1} - w_{n+1,m,k-1} - w_{n-1,m,k+1})/2 \\
&+ \lambda_1(\dot{u}_{n+1,m,k} - 2\dot{u}_{n,m,k} + \dot{u}_{n-1,m,k}) \\
&+ \lambda_2(\dot{u}_{n+1,m,k+1} + \dot{u}_{n-1,m,k-1} + \dot{u}_{n+1,m,k-1} + \dot{u}_{n-1,m,k+1} - 4\dot{u}_{n,m,k})/2 \\
&+ \lambda_2(\dot{u}_{n+1,m+1,k} + \dot{u}_{n-1,m-1,k} + \dot{u}_{n+1,m-1,k} + \dot{u}_{n-1,m+1,k} - 4\dot{u}_{n,m,k})/2 \\
&+ \lambda_2(\dot{v}_{n+1,m+1,k} + \dot{v}_{n-1,m-1,k} - \dot{v}_{n+1,m-1,k} - \dot{v}_{n-1,m+1,k})/2 \\
&+ \lambda_2(\dot{w}_{n+1,m,k+1} + \dot{w}_{n-1,m,k-1} - \dot{w}_{n+1,m,k-1} - \dot{w}_{n-1,m,k+1})/2,
\end{aligned} \quad (5)$$

$$\begin{aligned}
M\ddot{v}_{n,m,k} &= k_1(v_{n,m+1,k} - 2v_{n,m,k} + v_{n,m-1,k}) \\
&+ k_2(v_{n,m+1,k+1} + v_{n,m-1,k-1} + v_{n,m+1,k-1} + v_{n,m-1,k+1} - 4v_{n,m,k})/2 \\
&+ k_2(w_{n,m+1,k+1} + w_{n,m-1,k-1} - w_{n,m+1,k-1} - w_{n,m-1,k+1})/2 \\
&+ k_2(u_{n+1,m+1,k} + u_{n-1,m-1,k} - u_{n+1,m-1,k} - u_{n-1,m+1,k})/2 \\
&+ k_2(v_{n+1,m+1,k} + v_{n-1,m-1,k} + v_{n+1,m-1,k} + v_{n-1,m+1,k} - 4v_{n,m,k})/2 \\
&+ \lambda_1(\dot{v}_{n,m+1,k} - 2\dot{v}_{n,m,k} + \dot{v}_{n,m-1,k}) \\
&+ \lambda_2(\dot{v}_{n,m+1,k+1} + \dot{v}_{n,m-1,k-1} + \dot{v}_{n,m+1,k-1} + \dot{v}_{n,m-1,k+1} - 4\dot{v}_{n,m,k})/2 \\
&+ \lambda_2(\dot{w}_{n,m+1,k+1} + \dot{w}_{n,m-1,k-1} - \dot{w}_{n,m+1,k-1} - \dot{w}_{n,m-1,k+1})/2 \\
&+ \lambda_2(\dot{u}_{n+1,m+1,k} + \dot{u}_{n-1,m-1,k} - \dot{u}_{n+1,m-1,k} - \dot{u}_{n-1,m+1,k})/2 \\
&+ \lambda_2(\dot{v}_{n+1,m+1,k} + \dot{v}_{n-1,m-1,k} + \dot{v}_{n+1,m-1,k} + \dot{v}_{n-1,m+1,k} - 4\dot{v}_{n,m,k})/2,
\end{aligned}$$

$$M\ddot{w}_{n,m,k} = k_1(w_{n,m,k+1} - 2w_{n,m,k} + w_{n,m,k-1})$$
$$+ k_2(v_{n,m+1,k+1} + v_{n,m-1,k-1} - v_{n,m+1,k-1} - v_{n,m-1,k+1})/2$$
$$+ k_2(w_{n,m+1,k+1} + w_{n,m-1,k-1} + w_{n,m+1,k-1} + w_{n,m-1,k+1} - 4w_{n,m,k})/2$$
$$+ k_2(u_{n+1,m,k+1} + u_{n-1,m,k-1} - u_{n+1,m,k-1} - u_{n-1,m,k+1})/2$$
$$+ k_2(w_{n+1,m,k+1} + w_{n-1,m,k-1} + w_{n+1,m,k-1} + w_{n-1,m,k+1} - 4w_{n,m,k})/2$$
$$+ \lambda_1(\dot{w}_{n,m,k+1} - 2\dot{w}_{n,m,k} + \dot{w}_{n,m,k-1})$$
$$+ \lambda_2(\dot{v}_{n,m+1,k+1} + \dot{v}_{n,m-1,k-1} - \dot{v}_{n,m+1,k-1} - \dot{v}_{n,m-1,k+1})/2$$
$$+ \lambda_2(\dot{w}_{n,m+1,k+1} + \dot{w}_{n,m-1,k-1} + \dot{w}_{n,m+1,k-1} + \dot{w}_{n,m-1,k+1} - 4\dot{w}_{n,m,k})/2$$
$$+ \lambda_2(\dot{u}_{n+1,m,k+1} + \dot{u}_{n-1,m,k-1} - \dot{u}_{n+1,m,k-1} - \dot{u}_{n-1,m,k+1})/2$$
$$+ \lambda_2(\dot{w}_{n+1,m,k+1} + \dot{w}_{n-1,m,k-1} + \dot{w}_{n+1,m,k-1} + \dot{w}_{n-1,m,k+1} - 4\dot{w}_{n,m,k})/2.$$

Substituting (1) into (4), we obtain the equations of the motion of the mass with the coordinates ($n, m$) on the free surface $k = 0$ are as follows:

$$M\ddot{u}_{n,m,0} = k_1(u_{n+1,m,0} - 2u_{n,m,0} + u_{n-1,m,0})$$
$$+ k_2(u_{n+1,m+1,0} + u_{n-1,m-1,0} - 4u_{n,m,0} + u_{n+1,m-1,0} + u_{n-1,m+1,0})/2$$
$$+ k_2(u_{n-1,m,-1} - 2u_{n,m,0} + u_{n+1,m,-1})/2 + k_2(w_{n-1,m,-1} - w_{n+1,m,-1})/2$$
$$+ k_2(v_{n+1,m+1,0} + v_{n-1,m-1,0} - v_{n+1,m-1,0} - v_{n-1,m+1,0})/2$$
$$+ \lambda_1(\dot{u}_{n+1,m,0} - 2\dot{u}_{n,m,0} + \dot{u}_{n-1,m,0})$$
$$+ \lambda_2(\dot{u}_{n+1,m+1,0} + \dot{u}_{n-1,m-1,0} - 4\dot{u}_{n,m,0} + \dot{u}_{n+1,m-1,0} + \dot{u}_{n-1,m+1,0})/2$$
$$+ \lambda_2(\dot{u}_{n-1,m,-1} - 2\dot{u}_{n,m,0} + \dot{u}_{n+1,m,-1})/2 + \lambda_2(\dot{w}_{n-1,m,-1} - \dot{w}_{n+1,m,-1})/2$$
$$+ \lambda_2(\dot{v}_{n+1,m+1,0} + \dot{v}_{n-1,m-1,0} - \dot{v}_{n+1,m-1,0} - \dot{v}_{n-1,m+1,0})/2,$$

(6)

$$M\ddot{v}_{n,m,0} = k_1(v_{n,m+1,0} - 2v_{n,m,0} + v_{n,m-1,0})$$
$$+ k_2(v_{n+1,m+1,0} + v_{n-1,m-1,0} - 4v_{n,m,0} + v_{n+1,m-1,0} + v_{n-1,m+1,0})/2$$
$$+ k_2(v_{n,m-1,-1} - 2v_{n,m,0} + v_{n,m+1,-1})/2 + k_2(w_{n,m-1,-1} - w_{n,m+1,-1})/2$$
$$+ k_2(u_{n+1,m+1,0} + u_{n-1,m-1,0} - u_{n+1,m-1,0} - u_{n-1,m+1,0})/2$$
$$+ \lambda_1(\dot{v}_{n,m+1,0} - 2\dot{v}_{n,m,0} + \dot{v}_{n,m-1,0})$$
$$+ \lambda_2(\dot{v}_{n+1,m+1,0} + \dot{v}_{n-1,m-1,0} - 4\dot{v}_{n,m,0} + \dot{v}_{n+1,m-1,0} + \dot{v}_{n-1,m+1,0})/2$$
$$+ \lambda_2(\dot{v}_{n,m-1,-1} - 2\dot{v}_{n,m,0} + \dot{v}_{n,m+1,-1})/2 + k_2(\dot{w}_{n,m-1,-1} - \dot{w}_{n,m+1,-1})/2$$
$$+ \lambda_2(\dot{u}_{n+1,m+1,0} + \dot{u}_{n-1,m-1,0} - \dot{u}_{n+1,m-1,0} - \dot{u}_{n-1,m+1,0})/2,$$

$$M\ddot{w}_{n,m,0} = k_1(w_{n,m,-1} - w_{n,m,0}) + k_2(w_{n-1,m,-1} - 2w_{n,m,0} + w_{n+1,m,-1})/2$$
$$+ k_2(w_{n,m-1,-1} - 2w_{n,m,0} + w_{n,m+1,-1})/2 + k_2(u_{n-1,m,-1} - u_{n+1,m,-1})/2$$
$$+ k_2(v_{n,m-1,-1} - v_{n,m+1,-1})/2 + \lambda_1(\dot{w}_{n,m,-1} - \dot{w}_{n,m,0})$$
$$+ \lambda_2(\dot{w}_{n-1,m,-1} - 2\dot{w}_{n,m,0} + \dot{w}_{n+1,m,-1})/2$$
$$+ \lambda_2(\dot{w}_{n,m-1,-1} - 2\dot{w}_{n,m,0} + \dot{w}_{n,m+1,-1})/2 + \lambda_2(\dot{u}_{n-1,m,-1} - \dot{u}_{n+1,m,-1})/2$$
$$+ \lambda_2(\dot{v}_{n,m-1,-1} - \dot{v}_{n,m+1,-1})/2 + Q(t)\delta_{n0}\delta_{m0}.$$

Using the notation

$$\Lambda_{nn} f_{n,m,k} = f_{n+1,m,k} - 2f_{n,m,k} + f_{n-1,m,k},$$
$$\Phi_{nm} f_{n,m,k} = f_{n+1,m+1,k} + f_{n-1,m-1,k} - 4f_{n,m,k} + f_{n+1,m-1,k} + f_{n-1,m+1,k},$$
$$\Psi_{nm} f_{n,m,k} = f_{n+1,m+1,k} + f_{n-1,m-1,k} - f_{n+1,m-1,k} - f_{n-1,m+1,k},$$

we rewrite equations (5) in the following form:

$$M\ddot{u}_{n,m,k} = k_1 \Lambda_{nn} u_{n,m,k} + \frac{k_2[(\Phi_{nk} + \Phi_{nm})u_{n,m,k} + \Psi_{nm} v_{n,m,k} + \Psi_{nk} w_{n,m,k}]}{2}$$
$$+ \lambda_1 \Lambda_{nn} \dot{u}_{n,m,k} + \frac{\lambda_2[(\Phi_{nk} + \Phi_{nm})\dot{u}_{n,m,k} + \Psi_{nm} \dot{v}_{n,m,k} + \Psi_{nk} \dot{w}_{n,m,k}]}{2}, \quad (7)$$

$$M\ddot{v}_{n,m,k} = k_1 \Lambda_{mm} v_{n,m,k} + \frac{k_2[\Psi_{nm} u_{n,m,k} + (\Phi_{mk} + \Phi_{nm})v_{n,m,k} + \Psi_{mk} w_{n,m,k}]}{2}$$
$$+ \lambda_1 \Lambda_{mm} \dot{v}_{n,m,k} + \frac{\lambda_2[\Psi_{nm} \dot{u}_{n,m,k} + (\Phi_{mk} + \Phi_{nm})\dot{v}_{n,m,k} + \Psi_{mk} \dot{w}_{n,m,k}]}{2},$$

$$M\ddot{w}_{n,m,k} = k_1 \Lambda_{kk} w_{n,m,k} + \frac{k_2[\Psi_{nk} u_{n,m,k} + \Psi_{mk} v_{n,m,k} + (\Phi_{mk} + \Phi_{nk})w_{n,m,k}]}{2}$$
$$+ \lambda_1 \Lambda_{kk} \dot{w}_{n,m,k} + \frac{\lambda_2[\Psi_{nk} \dot{u}_{n,m,k} + \Psi_{mk} \dot{v}_{n,m,k} + (\Phi_{mk} + \Phi_{nk})\dot{w}_{n,m,k}]}{2}.$$

Using the notation

$$\Lambda_k^- f_{n,m,0} = f_{n+1,m,-1} - f_{n,m,0},$$
$$\Phi_{nk}^- f_{n,m,0} = f_{n-1,m,-1} - 2f_{n,m,0} + f_{n+1,m,-1}, \quad \Phi_{mk}^- f_{n,m,0} = f_{n,m-1,-1} - 2f_{n,m,0} + f_{n,m+1,-1},$$
$$\Psi_{nk}^- f_{n,m,0} = f_{n-1,m,-1} - f_{n+1,m,-1}, \quad \Psi_{mk}^- f_{n,m,0} = f_{n,m-1,-1} - f_{n,m+1,-1},$$

we rewrite equation (6) in the following form:

$$M\ddot{u}_{n,m,0} = k_1 \Lambda_{nn} u_{n,m,0} + \frac{k_2[(\Phi_{nk}^- + \Phi_{nm}^-)u_{n,m,0} + \Psi_{nm}^- v_{n,m,0} + \Psi_{nk}^- w_{n,m,0}]}{2}$$
$$+ \lambda_1 \Lambda_{nn} \dot{u}_{n,m,0} + \frac{\lambda_2[(\Phi_{nk}^- + \Phi_{nm}^-)\dot{u}_{n,m,0} + \Psi_{nm}^- \dot{v}_{n,m,0} + \Psi_{nk}^- \dot{w}_{n,m,0}]}{2}, \quad (8)$$

$$M\ddot{v}_{n,m,0} = k_1 \Lambda_{mm} v_{n,m,0} + \frac{k_2[\Psi_{nm}^- u_{n,m,0} + (\Phi_{mk}^- + \Phi_{nm}^-)v_{n,m,0} + \Psi_{mk}^- w_{n,m,0}]}{2}$$
$$+ \lambda_1 \Lambda_{mm} \dot{v}_{n,m,0} + \frac{\lambda_2[\Psi_{nm}^- \dot{u}_{n,m,0} + (\Phi_{mk}^- + \Phi_{nm}^-)\dot{v}_{n,m,0} + \Psi_{mk}^- \dot{w}_{n,m,0}]}{2},$$

$$M\ddot{w}_{n,m,0} = k_1 \Lambda_k^- w_{n,m,0} + \frac{k_2[\Psi_{nk}^- u_{n,m,0} + \Psi_{mk}^- v_{n,m,0} + (\Phi_{mk}^- + \Phi_{nk}^-)w_{n,m,0}]}{2}$$
$$+ \lambda_1 \Lambda_k^- \dot{w}_{n,m,0} + \frac{\lambda_2[\Psi_{nk}^- \dot{u}_{n,m,0} + \Psi_{mk}^- \dot{v}_{n,m,0} + (\Phi_{mk}^- + \Phi_{nk}^-)\dot{w}_{n,m,0}]}{2} + Q(t)\delta_{n0}\delta_{m0}.$$

Equations (7) and (8) describe the motion of the three-dimensional lattice of masses in Lamb's problem.

## 2. Results of numerical experiments. The case of a step load

In this section, we present the results of numerical calculations of disturbances in Lamb's problem for a lattice under the action of the vertical step load $Q(t) = H(t)$ applied at the point with coordinates $(0,0,0)$. All calculations, the results of which are presented in this section, are done for the case $l = M = 1$, $k_1 = 0.4$.

Figures 3–6 show the results of the calculations of the vertical $w$ and radial $u_r$ displacements and their velocities $\dot{w}$, $\dot{u}_r$ on the surface of the half-space at the moments of time $t = 50, 100, 150$ for $\lambda_1 = 0.1$, $\tau = 0.5$. Since the wave process is symmetric about the planes $Oxz$ and $Oyz$, figures 3–6 depict the results of the calculations for a single quadrant of the plane $Oxy$ only. The fronts of the Rayleigh waves are clearly visible on all plots shown on figures 3–6, while, on figure 5, the fronts of the longitudinal waves are also visible.

In order to analyze the spectral characteristics of the oscillations of the block medium, we calculate the spectral densities of $\dot{u}_r$ and $\dot{w}$ determined by the following formulas:

$$G(\dot{u}_r, \omega) = \left| \int_0^T \dot{u}_r(t) e^{-i\omega t} dt \right|, \qquad G(\dot{w}, \omega) = \left| \int_0^T \dot{w}(t) e^{-i\omega t} dt \right|.$$

The spectral densities are calculated numerically using the Fast Fourier Transform ($T = 150$).

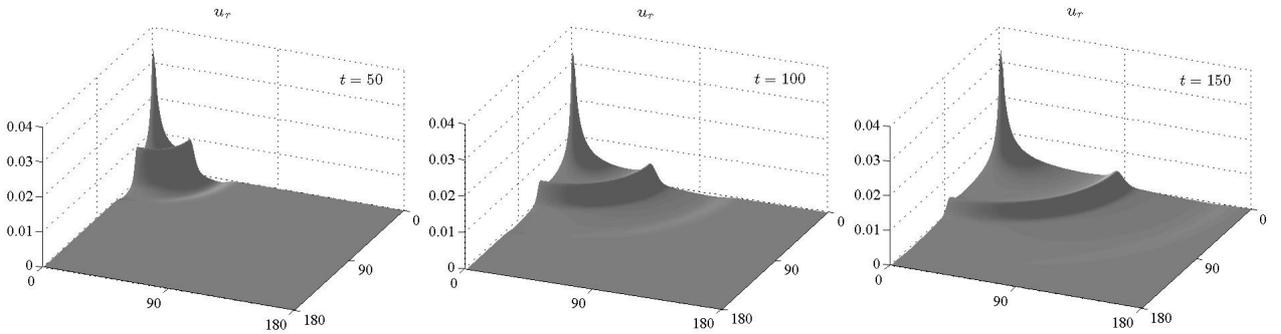

**Figure 3.** Radial displacement $u_r$ on the surface of the half-space at different moments of time.

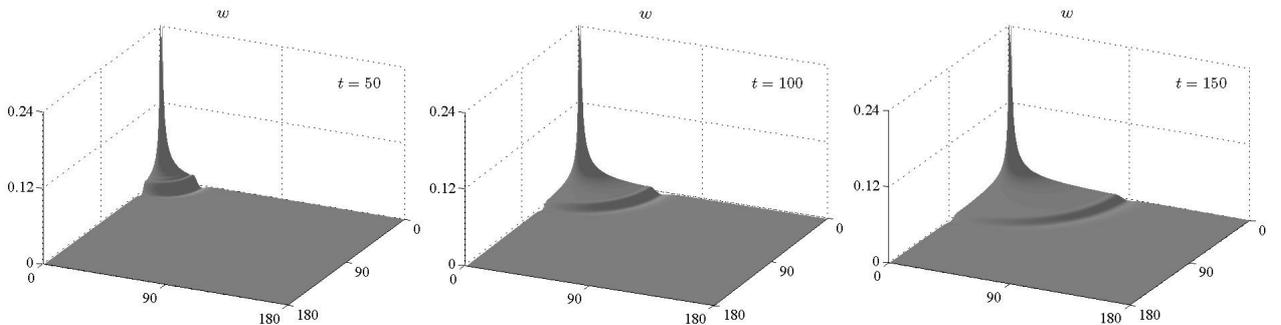

**Figure 4.** Vertical displacement $w$ on the surface of the half-space at different moments of time.

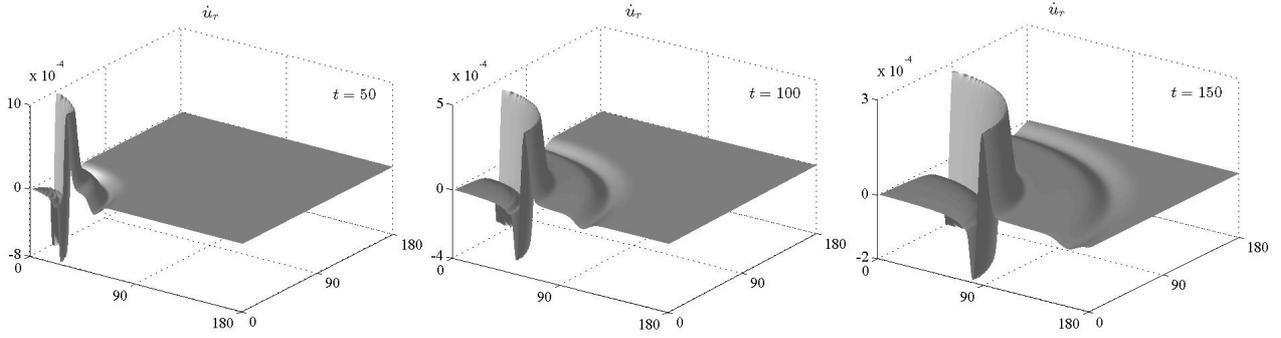

**Figure 5.** Radial velocity $\dot{u}_r$ of the displacement on the surface of the half-space at different moments of time.

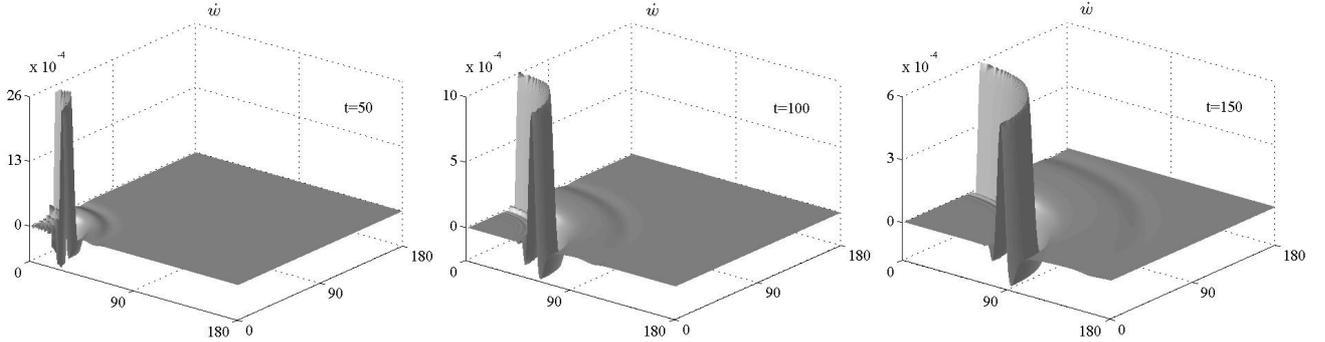

**Figure 6.** Vertical velocity $\dot{w}$ of the displacement on the surface of the half-space at different moments of time.

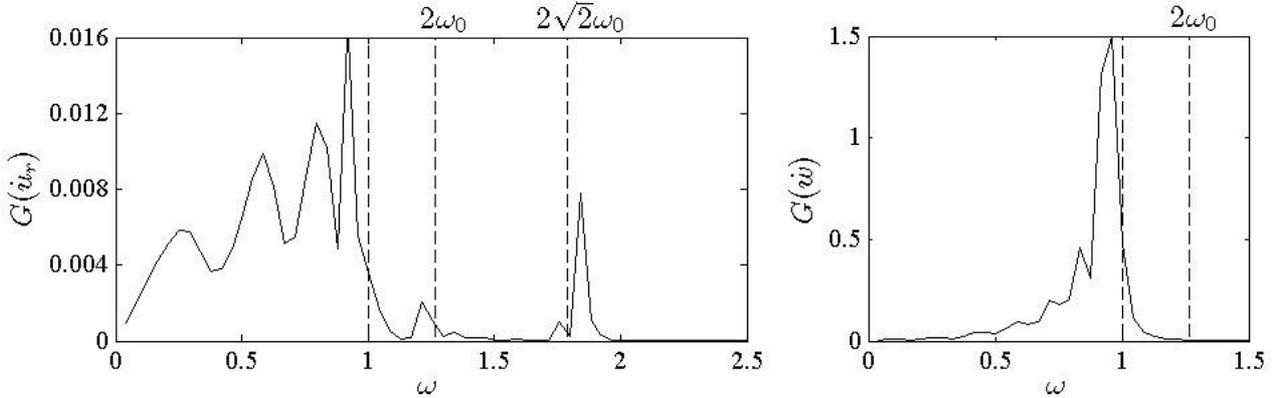

**Figure 7.** Spectral densities $G(\dot{u}_r, \omega)$ and $G(\dot{w}, \omega)$ of the velocities $\dot{u}_r$ and $\dot{w}$ of the displacements.

Figure 7 shows the plots of functions $G(\dot{u}_r, \omega)$ and $G(\dot{w}, \omega)$ on the surface of the half-space at the point $n = 20, m = 0$ under the action of the vertical step load for the case $\lambda_1 = 0$, $\tau = 0.25$. In figure 7, one vertical line corresponds to the frequency $\omega = \sqrt{2.5}\omega_0 = 1$, where $\omega_0 = \sqrt{k_1/M}$, of the short-wave perturbations in the Rayleigh wave on the surface of the block medium, while two other vertical lines correspond to the resonance frequencies $2\omega_0$ and $2\sqrt{2}\omega_0$ of the short-wave perturbations, propagating inside the block medium (see Section 3 of the article). From figure 7 we see that the spectra of the oscillations $\dot{w}$ arising in the block medium under the surface load are limited by the frequency of the surface waves: $\omega \leq 1$.

## 3. Results of numerical experiments. The case of a harmonic load

In this section, we present the results of numerical calculations of disturbances in Lamb's problem for a block medium under the action of the vertical harmonic load $Q(t) = H(t)\sin(\omega_* t)$ applied at the point with coordinates $(0,0,0)$. All calculations, the results of which are presented in this section, are done for the case $l = M = 1$, $k_1 = 0.4$, $\lambda_1 = 0$, $\tau = 0.5$.

Figure 8 shows the plots of the radial $\dot{u}_r$ and vertical $\dot{w}$ velocities of the displacements on the surface of the block half-space calculated at the point $n = 20, m = 0$. The red, black, and blue curves correspond to the frequencies $\omega_* = 0.5$, $\omega_* = 1$, and $\omega_* = 1.5$, respectively. In figure 8, we see the resonance growth of the oscillations under the load with the frequency $\omega_* = 1$. This indicates that $\omega_* = 1$ is the eigenfrequency of the surface waves. The vertical dashed lines in figure 8 correspond to the arrival time of the quasi-fronts of the longitudinal and Rayleigh waves at the point with the coordinates $(n,0,0)$: $t_p = n/c_p$, $t_R = n/c_R$, where $c_p, c_R$ are defined by the formulas (3.2) and (3.3) of the article.

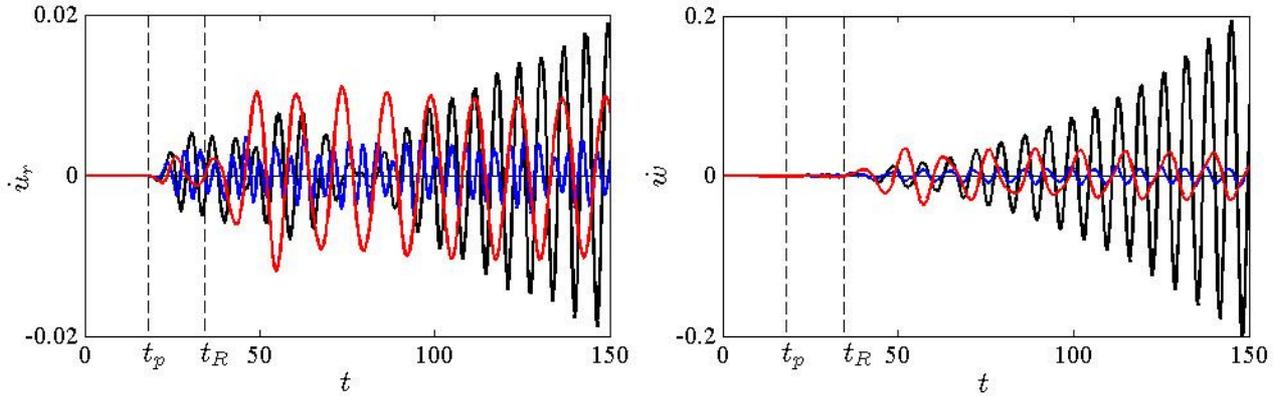

**Figure 8.** Radial $\dot{u}_r$ and vertical $\dot{w}$ velocities of the displacements on the surface of the block half-space calculated at the point $n = 20, m = 0$.

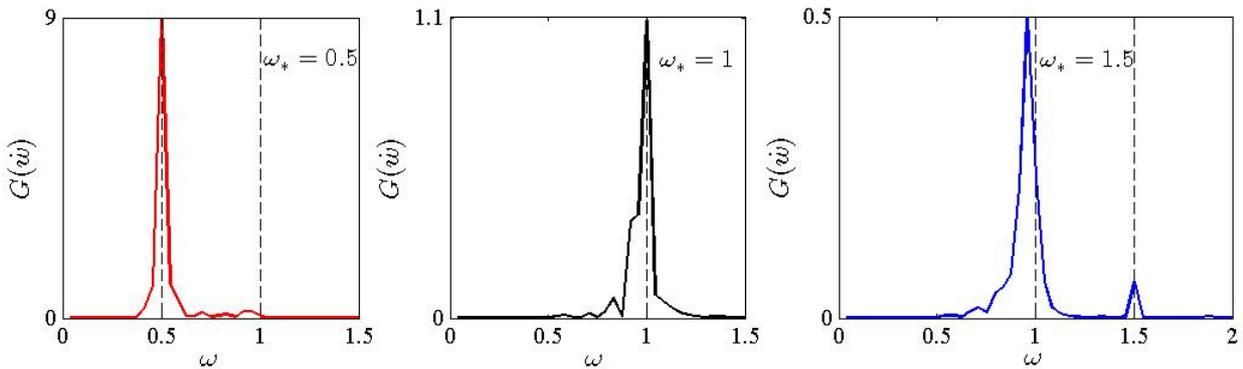

**Figure 9.** Spectral densities $G(\dot{w},\omega)$ of the velocities $\dot{w}$ of the displacement, which are shown in figure 8.

Figure 9 depicts the plots of the function $G(\dot{w},\omega)$ for $\dot{w}$, which are shown in figure 8. Figures 8 and 9 show that the block medium practically does not pass perturbations with the frequency $\omega_* > 1$ while perturbations with the frequency $\omega_* < 1$ pass through the medium easily.

## 4. Results of numerical experiments. The case of a Gauss pulse, Ricker function and tapered sinusoid

In this section, we present the results of numerical calculations of disturbances in Lamb's problem for a block medium under the action of the vertical loads defined by the following formulas:

(i) $Q(t) = \dfrac{12.5}{\sqrt{2\pi}\sigma} e^{-\dfrac{(t-t_0)^2}{2\sigma^2}}$, where $t_0 = 20$, $\sigma = 5$ (the Gauss pulse),

(ii) $Q(t) = \left(1 - \dfrac{(t-t_0)^2}{\sigma^2}\right) e^{-\dfrac{(t-t_0)^2}{2\sigma^2}}$, where $t_0 = 20$, $\sigma = 5$ (the Ricker function),

(iii) $Q(t) = \sin(\omega_* t)\sin\left(\dfrac{\omega_* t}{9}\right)$, where $\omega_* = \dfrac{9\pi}{40} \approx 0.707$ (the tapered sinusoid).

The loads are applied at the point with coordinates $(0,0,0)$. All calculations, the results of which are presented in this section, are done for the case $l = M = 1$, $k_1 = 0.4$, $\lambda_1 = 0$, $\tau = 0.5$.

Figure 10 shows the plots of the loads $Q(t)$ defined by (i)–(iii). In figure 10, the blue, red, and black curves correspond to the Gauss pulse, Ricker function, and tapered sinusoid, respectively.

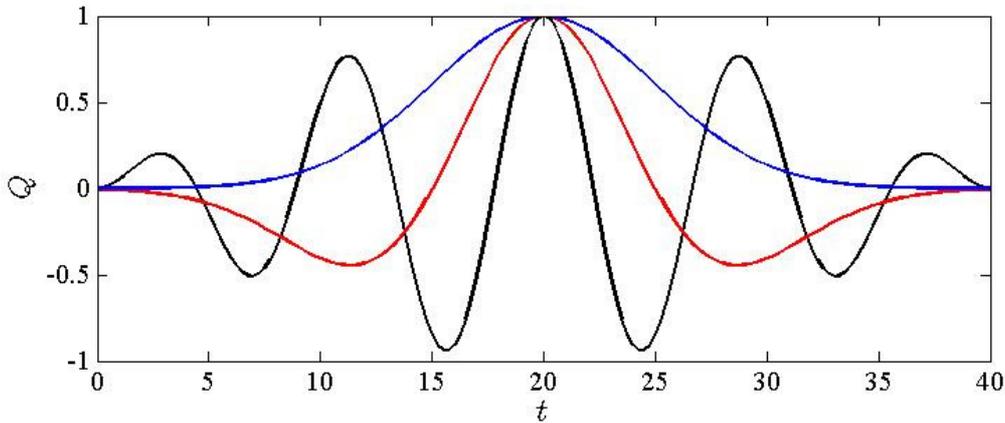

**Figure 10.** Loads (i)–(iii) versus time. The blue, red, and black curves correspond to the Gauss pulse, Ricker function, and tapered sinusoid, respectively.

Figures 11–13 show the plots of the radial $\dot{u}_r$ and vertical $\dot{w}$ velocities of the displacements on the surface of the half-space calculated at the point $n = 30, m = 0$ and the plots of the spectral density $G(\dot{w},\omega)$ for the cases of the Gauss pulse, Ricker function, and tapered sinusoid. The vertical dashed lines in figures 11–13 correspond to the arrival time of the quasi-fronts of the longitudinal

and Rayleigh waves at the point with the coordinates $(n,0,0)$: $t_p = n/c_p$, $t_R = n/c_R$, where $c_p, c_R$ are defined by the formulas (3.2) and (3.3) of the article.

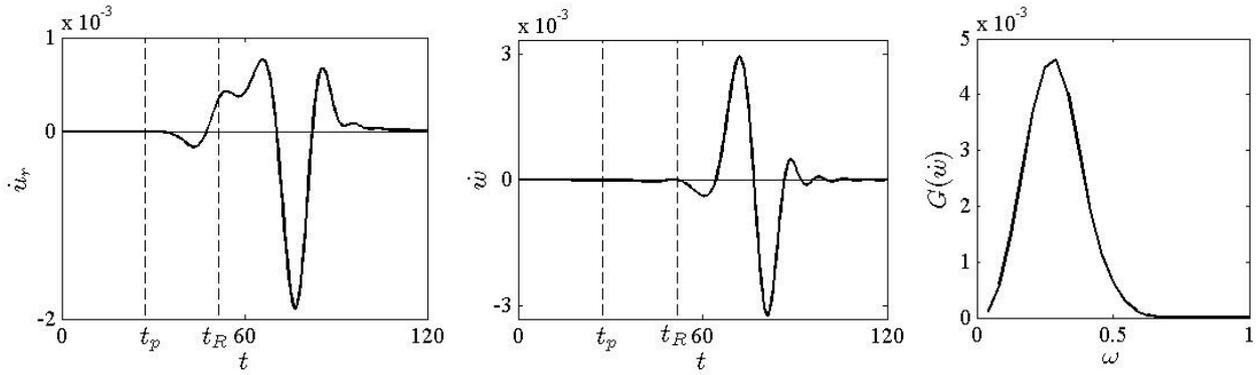

**Figure 11.** Case of the Gauss pulse: radial $\dot{u}_r$ and vertical $\dot{w}$ velocities of the displacements on the surface of the block half-space calculated at the point $n = 30, m = 0$ and the spectral density $G(\dot{w}, \omega)$.

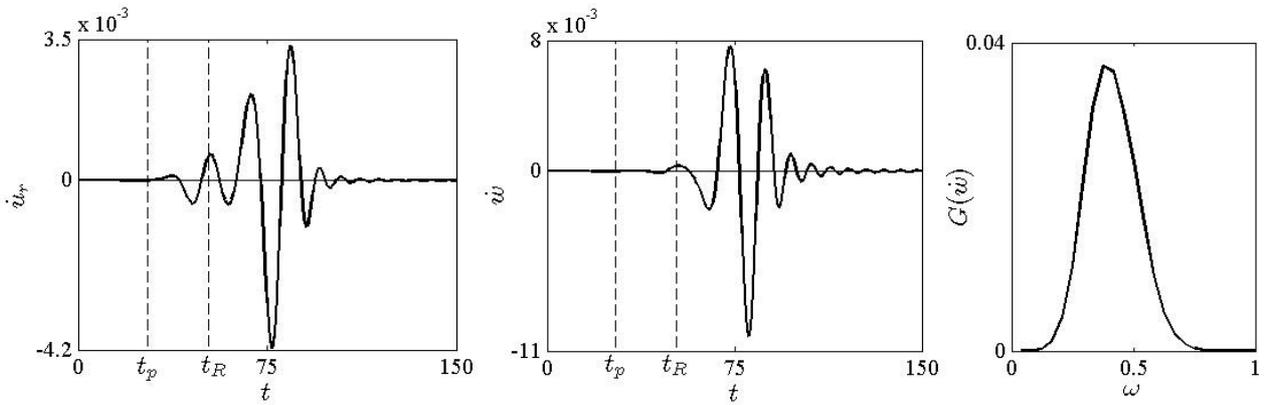

**Figure 12.** Case of the Ricker function: radial $\dot{u}_r$ and vertical $\dot{w}$ velocities of the displacements on the surface of the block half-space calculated at the point $n = 30, m = 0$ and the spectral density $G(\dot{w}, \omega)$.

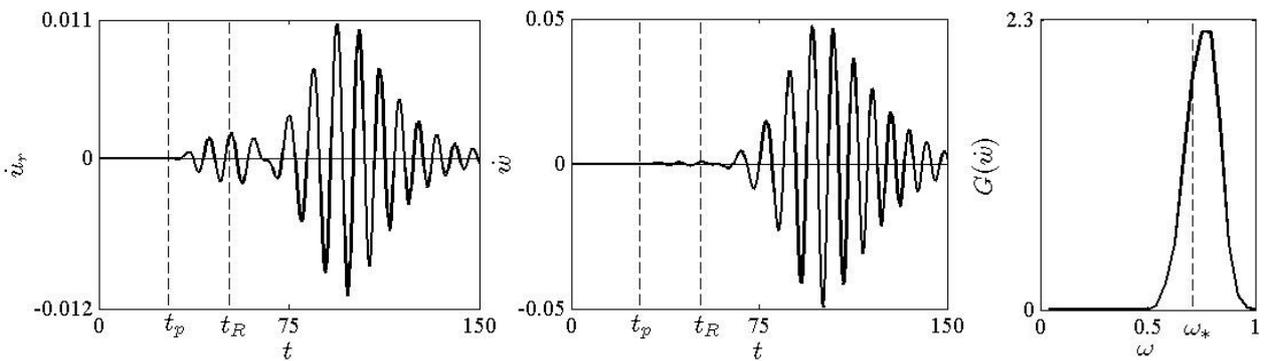

**Figure 13.** Case of the tapered sinusoid: radial $\dot{u}_r$ and vertical $\dot{w}$ velocities of the displacements on the surface of the block half-space calculated at the point $n = 30, m = 0$ and the spectral density $G(\dot{w}, \omega)$.

Summarizing, we can say that the results of the calculations presented in Sections 3 and 4 of the electronic supplementary material show that the high-frequency oscillations generated by a step load are caused by dispersion inherent to the block medium, not by spurious oscillations induced by numerical dispersion.